\documentclass[aps,preprint,showpacs,eqsecnum,superscriptaddress,notitlepage,floatfix, pre]{revtex4-1}
\usepackage{graphicx}
\usepackage{dcolumn}
\usepackage{bm}
\usepackage{amsmath}
\usepackage{amssymb}
\usepackage{verbatim}
\usepackage{color}
\usepackage{hyperref}
\usepackage{tikz}
\usetikzlibrary{arrows}
\usetikzlibrary{positioning,arrows,patterns}
\usetikzlibrary{decorations.markings}
\usetikzlibrary{calc}
\usepackage{xparse}
\usepackage{feynmf}
\NewDocumentCommand\semiloop{O{black}mmmO{}O{above}}
{
\draw[#1] let \p1 = ($(#3)-(#2)$) in (#3) arc (#4:({#4+180}):({0.5*veclen(\x1,\y1)})node[midway, #6] {#5};)
}
\NewDocumentCommand\negsemiloop{O{black}mmmO{}O{above}}
{%
\draw[#1] let \p1 = ($(#3)-(#2)$) in (#3) arc (#4:({#4-180}):({0.5*veclen(\x1,\y1)})node[midway, #6] {#5};)
}

\tikzset{
  photon/.style={decorate, decoration={snake}, draw=black},
  fermion/.style={draw=black, postaction={decorate},decoration={markings,mark=at position .55 with {\arrow{>}}}},
  vertex/.style={draw,shape=circle,fill=black,minimum size=3pt,inner sep=0pt},
}

\begin{document}
\author{Si-Wei Qiu}
\author{Carson C. Chow}
\affiliation{Mathematical Biology Section, LBM, NIDDK, NIH}
\title{Finite size effects for spiking neural networks with spatially dependent coupling}
\date{\today}
\begin{abstract}
We study finite-size fluctuations in a network of spiking deterministic neurons coupled with non-uniform synaptic coupling. We generalize a previously developed theory of finite size effects 
for globally coupled neurons with a uniform coupling function. In the uniform coupling case, mean field theory is well defined by averaging over the network as the number of neurons in the network goes to infinity. However, for nonuniform coupling it is no longer possible to average over the entire network if we are interested in fluctuations at a particular location within the network. We show that if the coupling function approaches a continuous function in the infinite system size limit then an average over a local neighborhood can be defined such that mean field theory is well defined for a spatially dependent field. We then use a path integral formalism to derive a perturbation expansion in the inverse system size around the mean field limit for the covariance of the input to a neuron (synaptic drive) and firing rate fluctuations due to dynamical deterministic finite-size effects.
\end{abstract}
\maketitle

\section{Introduction}
The dynamics of neural networks have traditionally been studied in the limit of very large numbers of neurons, where mean field theory can be applied, e.g.~\cite{Amari1977, Wilson, Hopfield2554, COHEN1987288, Mirollo1990, Treves1993, Abbott1993, Brunel1999, Brunel2002, Duane2006} or for a small number of neurons, where traditional dynamical systems approaches can be used, e.g.~\cite{Bard1996,Jones2006,maran2008}.  The intermediate regime of large but finite numbers of neurons can have interesting properties that are independent of the small and infinite system limits~\cite{2007MichaelPRE, 2007MichaelPRL, finite2013,Touboul2011,Helias2013,Helias2016,Lang2017,DumontFiniteSize2017}.  However, these previous works have not fully explored fluctuations due to finite-size effects at specific locations within the network when all the  neurons receive nonhomogeneous input from other neurons because of nonuniform coupling.
Here, we consider finite-size effects in a network of spiking neurons with nonuniform synaptic coupling.   Previously~\cite{2007MichaelPRE, 2007MichaelPRL, finite2013}, a perturbation expansion in the inverse network neuron number had been developed for networks with global spatially uniform coupling and we generalize that theory to include nonuniform coupling.  We first show that mean field theory in the infinite nonuniform system limit can be realized in a single network if a spatial metric can be imposed on the network and the coupling function is a continuous function of this distance measure.  We then analyze finite-size fluctuations around such mean field solutions using a path integral formalism to derive a perturbation expansion in the inverse network neuron number for the spatially dependent covariance function for the synaptic drive and spatially dependent neuron firing rate.

\section{Coupled neuron model}
Consider a network of $N$ theta neurons (phase reduction of quadratic integrate-and-fire neurons~\cite{Bard1996}) on a one dimensional periodic domain of size $L$ although the theory can be applied to any domain.  The network obeys the following deterministic {\it microscopic} equations:
\begin{align}
\dot{\theta_i}&=1-\cos\theta_i+(I_i+u_i(t))(1+\cos\theta_i)\\
u_i &= \frac{L}{N}\sum_{j=1}^Nw_{ij} s_j\\
\dot{s}_j&=-\beta s_j+\beta \sum_l \delta(t-t_j^l)\label{eq:drivea}
\end{align}
where $\theta_i$ is the phase of neuron $i$, $u_i$ is the synaptic drive to neuron $i$, $I_i$ is the external input to neuron $i$, $\beta$ is the decay constant of the synaptic drive, $s_j$ is the time dependent synaptic input from neuron $j$, and $t_j^l$ represents the spike times when the phase of neuron $j$ crosses $\pi$.   $s_j$ rises instantaneously when neuron $j$ spikes and relaxes to zero with a time constant of $1/\beta$.  The synaptic drive represents the total time dependent synaptic input where the contribution from each neuron is weighted by the synaptic coupling function $w_{ij}$ (a real $N\times N$ matrix).  When $I_i + u_i >0$, the neuron receives suprathreshold input and $\theta_i$ will progress in time. When it passes $\pi$, the neuron is said to spike.  When $I_i +u_i <0$ the neuron receives subthreshold input and the phase will approach a fixed point.  The theta neuron is the normal form of a Type I spiking neuron near the bifurcation point to firing~\cite{Bard1996}.  By linearity, the synaptic drive obeys the more convenient form of
\begin{align}
\dot{u}_i&=-\beta u_i+\beta \frac{L}{N}\sum_{j=1}^Nw_{ij}\sum_l \delta(t-t_j^l)\label{eq:drive1}
\end{align}


We define an empirical density
\begin{align}
\eta_j(\theta,t) = \delta(\theta-\theta_j(t))
\end{align}
that assigns a point mass to the phase of each neuron in the network.  Hence, we can write the sum of a spike train as $\sum_l\delta(t-t_j^l) = \eta_j(\pi,t)\dot{\theta}_j|_{\theta_j=\pi}$.  For the theta model, $\dot{\theta}_j|_{\theta_j=\pi}=2$ and thus 
we can then rewrite (\ref{eq:drive1}) as
\begin{align}
\dot{u}_i=-\beta u_i+2\beta \frac{L}{N}\sum_{j=1}^Nw_{ij}\eta_j(\pi,t)
\label{eq:drive}
\end{align}

Neuron number is conserved so the neuron density formally obeys a conservation (Klimontovich) equation~\cite{finite2013}:
\begin{eqnarray}
\partial_t \eta_i(\theta,t) + \partial_\theta F_i(\theta,u_i)\eta_i(\theta,t)=0
\label{eq:klimontovich}
\end{eqnarray}
where $F_i(\theta,u_i)=1-\cos\theta+(1+\cos\theta)(I_i+ u_i)$.  The Klimontovich equation together with (\ref{eq:drive}) fully describes the system.  However, it is only a formal definition since $\eta$ is not in general differentiable.  In the following, we develop a method to regularize the Klimontovich equation so that desired quantities can be calculated.

\section{Mean field theory}

The Klimontovich equation (\ref{eq:klimontovich}) only exists in a weak sense. We can regularize it by taking a suitable average over an ensemble of initial conditions:
\begin{eqnarray}
\partial_t \langle \eta_i(\theta,t)\rangle + \partial_\theta \langle F_i(\theta,u_i)\eta_i(\theta,t) \rangle=0
\end{eqnarray}
This equation is not closed because it involves covariances such as $\langle \eta\eta \rangle$, which in turn depend on higher order cumulants in a BBGKY hierarchy~\cite{2007MichaelPRE, 2007MichaelPRL, finite2013}.  This hierarchy can be rendered tractable if we can truncate it.
Mean field theory truncates the hierarchy at first order by assuming that all cumulants beyond the first are zero so we can write
\begin{eqnarray}
\partial_t \rho_i(\theta,t)+ \partial_\theta F_i(\theta, a_i)\rho_i(\theta,t)=0
\end{eqnarray}
where $a_i=\langle u_i\rangle$ and $\rho_i=\langle \eta_i \rangle$.  The full set of closed mean field equations are given by
\begin{eqnarray}
\partial_t  \rho_i(\theta,t) + \partial_\theta \left[1-\cos\theta+(I_i+ a_i)(1+\cos\theta)\right]\rho_i(\theta,t)=0\nonumber\\
\dot{a}_i=-\beta a_i+2\beta\frac{L}{N}\sum_jw_{ij}\rho_j(\pi,t)
\label{eqn:meanfield}
\end{eqnarray}

Although we can always write down the mean field equations (\ref{eqn:meanfield}),  it is not clear that a given network would obey it in the infinite $N$ limit.  In previous work~\cite{Desai1978,Mirollo1990,finite2013},  it was shown that mean field theory applies to a network of coupled oscillators with uniform coupling in the infinite $N$ limit.
However, it is not known when or if mean field theory applies for nonuniform coupling.


To see this, consider first the stationary system 
\begin{align}
 &\partial_\theta \left[1-\cos\theta+(I_i + u_i)(1+\cos\theta)\right]\eta_i(\theta)=0\label{eq:etastat}\\
 &u_i=2\frac{L}{N}\sum_jw_{ij} \eta_j(\pi)\label{eq:drivestat}
\end{align}
with uniform coupling, $w_{ij}=w$, and uniform external input, $I_i = I$.  If the neurons are initialized with random phases and remain asynchronous then we can suppose that in the limit of $N\rightarrow \infty$ the quantity
\begin{align}
\rho(\pi) =  \frac{L}{N}\sum_j \eta_j(\pi)
\end{align}
converges to an invariant quantity~\cite{Desai1978,Mirollo1990,finite2013}. This then implies that
$u_i=2w \rho \equiv a$ is also a constant.
Thus each neuron will have identical inputs so
if we apply the network averaging operator $\frac{L}{N}\sum_ {i=1}^N $  to (\ref{eq:etastat})  we obtain
\begin{align}
& \partial_\theta \left[1-\cos\theta+(I + a)(1+\cos\theta)\right]\rho(\theta)=0\\
& a = 2 w \rho(\pi)
\end{align}
Covariances vanish and mean field theory is realized in the infinite network limit.
Given that the drive equation (\ref{eq:drive}) is linear, the time dependent mean field theory will similarly hold in the large $N$ limit.

In the case where $w_{ij}$ is not uniform, covariances are not guaranteed to vanish and an infinite network need not obey mean field theory. Our goal is to find conditions such that mean field theory applies.  Again, consider the stationary equations (\ref{eq:etastat}) and (\ref{eq:drivestat}).  Now, instead of averaging over the entire domain, take a local interval around $j$, $[j-cN/2,j+cN/2]$, where $c<1$ is a constant that can depend on $N$ and we map $j-cN/2 < 1$ to $N+j - cN/2$ and $j+cN/2>N$ to $j+cN/2-N$.
We want to express our mean field equation in terms of the locally averaged empirical density
\begin{align}
\rho_j=\frac{1}{cN}\sum_{k=j-\frac{cN}{2}}^{j+\frac{cN}{2}} \eta_k\label{eq:empiricaldens}
\end{align}
 If $cN\rightarrow \infty$ for $N\rightarrow\infty$ then it is feasible that the local empirical density can be invariant (to random initial conditions) and correlations can vanish; we seek conditions on the coupling for which this is true.
Inserting (\ref{eq:drivestat}) into (\ref{eq:etastat}), and taking the local average  yields
\begin{align}
 \frac{1}{cN}\sum_{i=k-\frac{cN}{2}}^{k+\frac{cN}{2}} \partial_\theta \left[1-\cos\theta+
 \left(I_i +2\frac{L}{N}\sum_{j=1}^Nw_{ij}\eta_{j}(\pi)\right)(1+\cos\theta)\right]\eta_i(\theta)=0
 \label{eq:etastat3}
 \end{align}
We immediately see that correlations can arise from the sums over the product of $\eta_j(\pi)\eta_i(\theta)$. Consider the identity $\sum_{j=1}^{N}w_{ij}\eta_j(\pi) =\sum_{j=1}^{N}(cN)^{-1}\sum_{l=j-\frac{cN}{2}}^{j+\frac{cN}{2}}w_{il}\eta_l(\pi)$, which is exact for periodic boundary conditions.  For nonperiodic boundary conditions there will be an edge contribution but this should be negligible in the large network limit.
Using this summation identity, we can rewrite the sum as
\begin{align}
\sum_{j=1}^{N}\frac{1}{cN}\sum_{i=k-\frac{cN}{2}}^{k+\frac{cN}{2}}w_{ij}\eta_j(\pi)\eta_i(\theta)
=\sum_{j=1}^{N}\left(w_{kj}\rho_j(\pi)\rho_k(\theta)+ R_{jk}\right)
\end{align}
where the remainder
\begin{align}
R_{jk} &= \frac{1}{(cN)^2}\sum_{l=j-\frac{cN}{2}}^{j+\frac{cN}{2}}\sum_{i=k-\frac{cN}{2}}^{k+\frac{cN}{2}}(w_{ij}-w_{kj})\rho_j(\pi)\eta_i(\theta)\nonumber\\
&+\frac{1}{(cN)^2}\sum_{l=j-\frac{cN}{2}}^{j+\frac{cN}{2}}\sum_{i=k-\frac{cN}{2}}^{k+\frac{cN}{2}}(w_{il}-w_{ij})\eta_l(\pi)\eta_i(\theta)
\end{align}
carries the correlations. Mean field theory is valid in the $N\rightarrow\infty$ limit if $R_{jk}$ vanishes.  Its magnitude obeys 
\begin{align}
|R_{jk}|
\leq&\bigg|\frac{1}{(cN)^2}\sum_{l=j-\frac{cN}{2}}^{j+\frac{cN}{2}}\sum_{i=k-\frac{cN}{2}}^{k+\frac{cN}{2}}(w_{ij}-w_{kj})\rho_j(\pi)\eta_i(\theta)\bigg|+\bigg|\frac{1}{(cN)^2}\sum_{l=j-\frac{cN}{2}}^{j+\frac{cN}{2}}\sum_{i=k-\frac{cN}{2}}^{k+\frac{cN}{2}}(w_{il}-w_{ij})\eta_l(\pi)\eta_i(\theta)\bigg|\nonumber\\
\leq&\rho_j(\pi)\bigg(\frac{1}{cN}\sum_{i=k-\frac{cN}{2}}^{k+\frac{cN}{2}}\eta_i(\theta)\bigg)\sup_{i\in(k-\frac{cN}{2},k+\frac{cN}{2})}|w_{ij}-w_{kj}|\nonumber\\
&+\bigg(\frac{1}{(cN)^2}\sum_{l=j-\frac{cN}{2}}^{j+\frac{cN}{2}}\sum_{i=k-\frac{cN}{2}}^{k+\frac{cN}{2}}(\eta_l(\pi)\eta_i(\theta))\bigg)\sup_{i\in(k-\frac{cN}{2},k+\frac{cN}{2}),l\in(j-\frac{cN}{2}, j+\frac{cN}{2})}|w_{il}-w_{ij}|\label{eq:bd1}
\end{align}
since the density is nonnegative.  

Applying (\ref{eq:empiricaldens}) then leads to
\begin{align}
|R|\leq&\rho_j(\pi)\rho_k(\theta)\left(\sup_{i\in(k-\frac{cN}{2},k+\frac{cN}{2})}|w_{ij}-w_{kj}|+\sup_{i\in(k-\frac{cN}{2},k+\frac{cN}{2}),l\in(j-\frac{cN}{2}, j+\frac{cN}{2})}|w_{il}-w_{ij}|\right)\label{eq:bd2}
\end{align}
We introduce a distance measure $z=i L/N$, $z'=jL/N$ ,$z'' = kL/N$, $z''' = lL/N$ and write $\rho_i(\theta)|_{i=zN/L}=\rho(z,\theta)$ and $w_{ij}|_{i=zN/L,j=z'N/L}=w(z,z')$.  Then
\begin{align}
|R|\leq\rho(z',\pi)\rho(z'',\theta)\left(
 \sup_{z\in[z''-cL/2,z''+cL/2]}|w(z,z')-w(z'',z')|\right.\\
\left.+
 \sup_{z\in[z''-cL/2,z''+cL/2], z'''\in[z'-cL/2,z'+cL/2]}
|w(z,z''')-w(z,z')|\right)\label{eq:bd2}
\end{align}
Hence, if we set $c=N^{-\alpha}$, $0<\alpha<1$, then as $N\rightarrow\infty$, the number of neurons in the local neighborhood $cN$ approaches infinity as $N^{1-\alpha}$ while $c\rightarrow 0$.    Then, $|R|\rightarrow 0$ as $N \rightarrow \infty$ if $\lim_{z\rightarrow z''} w(z,z') -w(z'',z')=0$ and $\lim_{z'''\rightarrow z'} w(z,z''') -w(z,z') =0$, i.e.
$w_{ij}$ is a continuous function in both indices.
A similar argument shows that 
\begin{align}
 \frac{1}{cN}\sum_{i=k-\frac{cN}{2}}^{k+\frac{cN}{2}}( I_i -I_k)\eta_i(\theta) \rightarrow 0
 \end{align}
if  $I_i$ approaches a continuous function in index $i$ in the infinite $N$ limit.  Then (\ref{eq:etastat}) and (\ref{eq:drivestat}) can be written as
\begin{align}
 &\partial_\theta \left[1-\cos\theta+
 \left(I_k +a_k\right)(1+\cos\theta)\right]\rho_k(\theta)=0
 \label{eq:etastat2}\\
& a_k = 2 \frac{L}{N}\sum_j w_{kj} \rho_j(\pi)
\label{eq:drivestat2}
 \end{align}
Equations (\ref{eq:etastat2}) and (\ref{eq:drivestat2}) form a mean field theory that is realized in a nonuniform coupled network in the infinite size limit as long as the input and coupling function are continuous functions.  By linearity, the time dependent mean field theory should equally apply if the external input and the coupling are continuous functions of the indices.

In the $N\rightarrow\infty$ limit, setting $i\rightarrow z N/L$, $a_i(t)\rightarrow a(z,t)$,  $\rho_i(\theta,t)\rightarrow \rho(z,\theta,t)$, $I_i\rightarrow I(z)$ is continuous, $\sum_i \rightarrow (N/L)\int_\Omega dz$, and $w_{ij} \rightarrow w(z,z')$ is continuous, we can write mean field theory in continuum form as
\begin{eqnarray}
\partial_t  \rho(z,\theta,t) + \partial_\theta \left[1-\cos\theta+(I(z)+ a(z,t))(1+\cos\theta)\right]\rho(z,\theta,t)=0\nonumber\\
\partial_t{a}(z,t)=-\beta a(z,t)+2\beta\int w(z,z')\rho(z',\pi,t) dz'
\label{eqn:mfcont}
\end{eqnarray}

\begin{figure}[t!]
\begin{minipage}{.32\textwidth}
\includegraphics[width=1\linewidth, height=0.2\textheight]{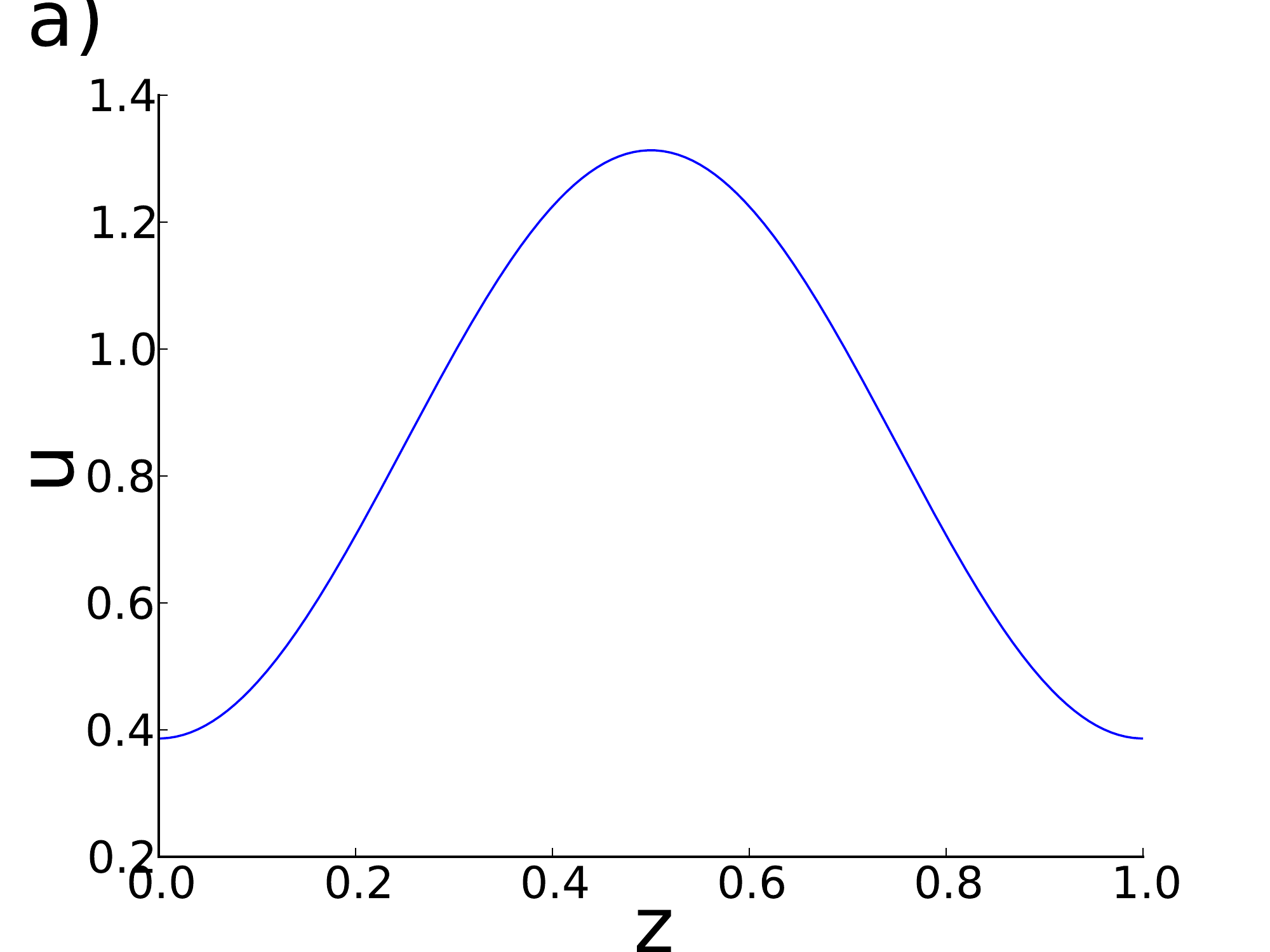} 
\end{minipage}
\begin{minipage}{.32\textwidth}
\includegraphics[width=1\linewidth, height=0.2\textheight]{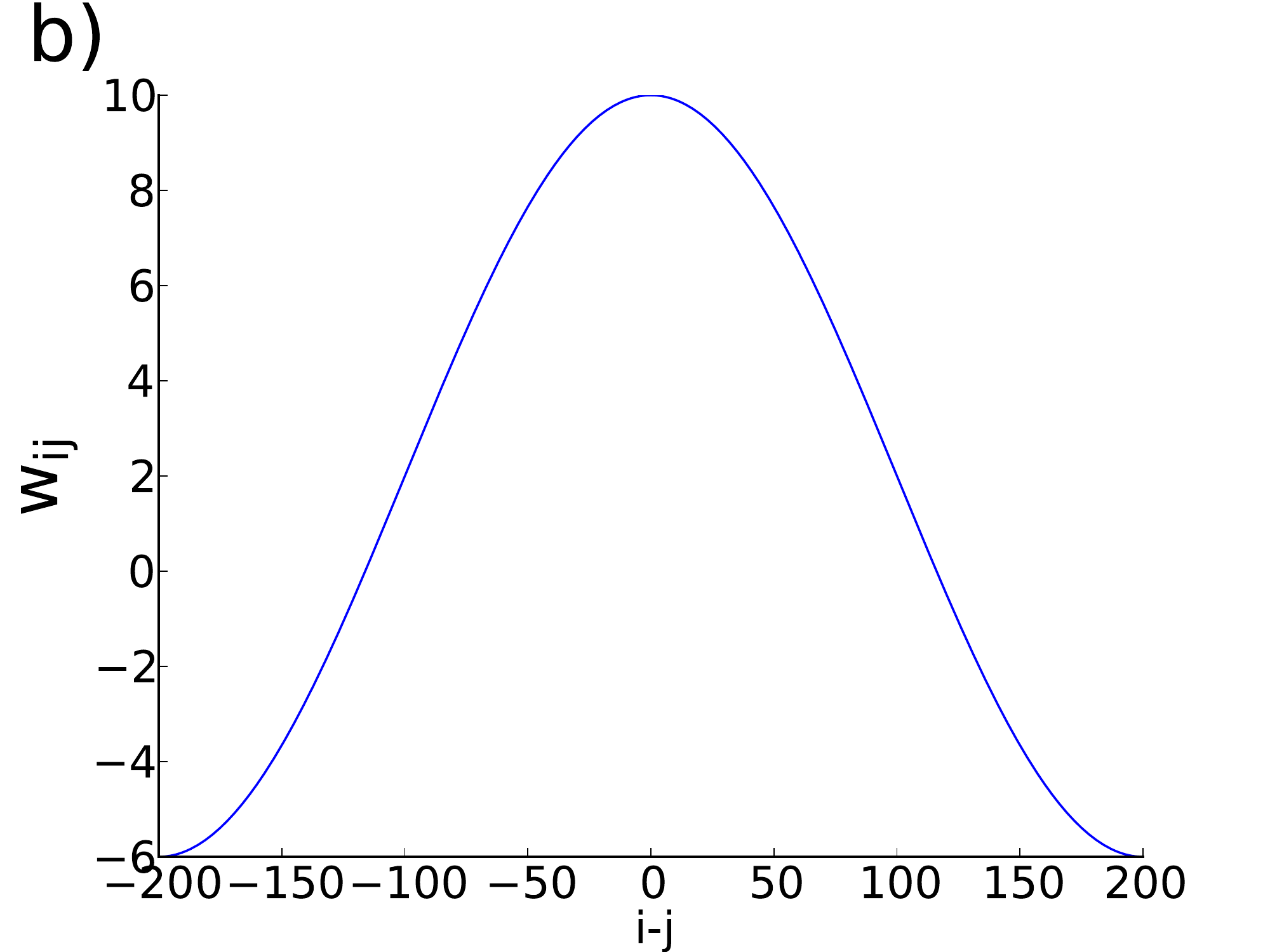} 
\end{minipage}
\begin{minipage}{.32\textwidth}
\includegraphics[width=1\linewidth, height=0.2\textheight]{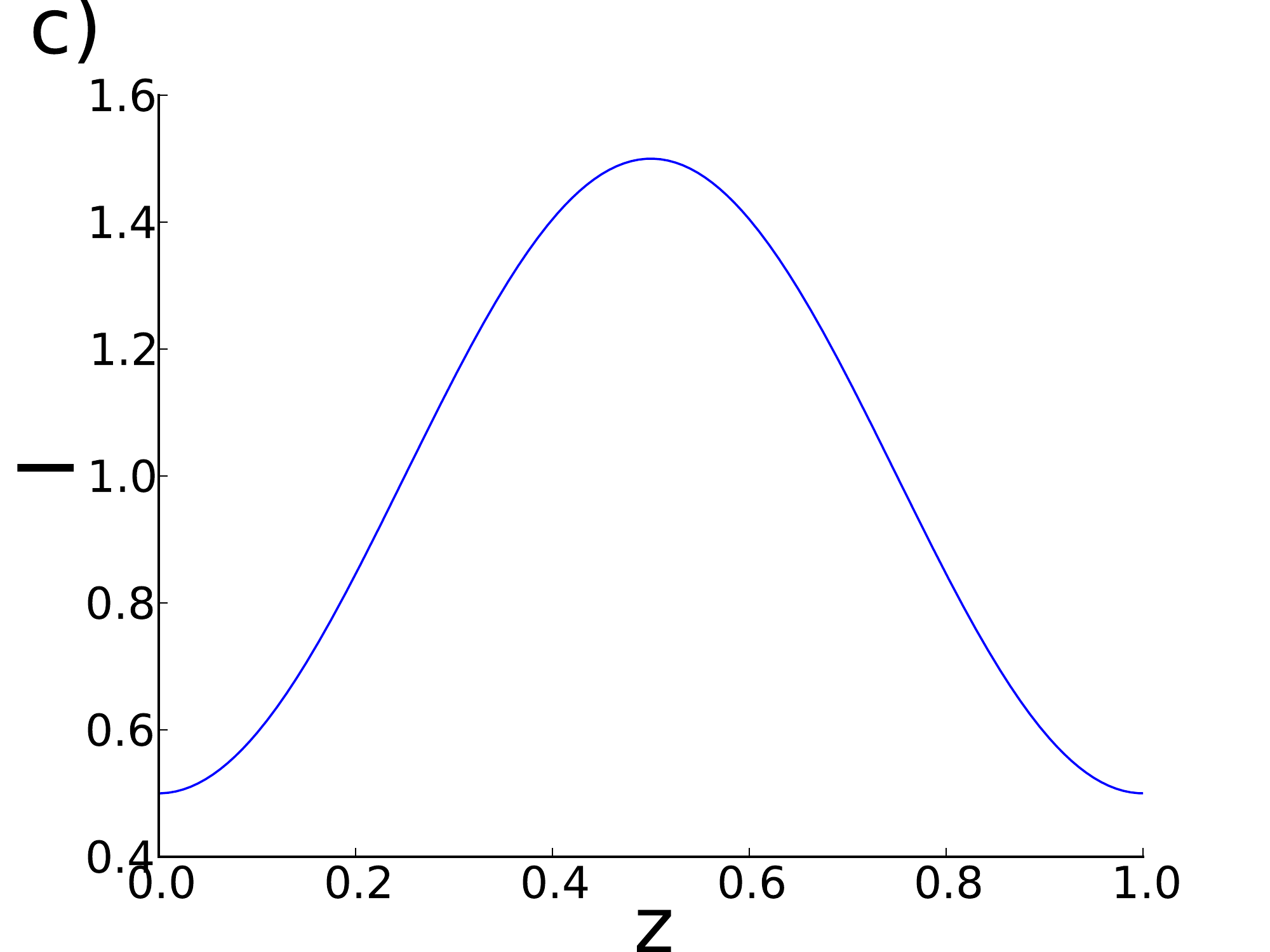} 
\end{minipage}
\caption{a) Mean field theory synaptic drive for b)
connectivity weight $w(z)=-J_0+J_2 \cos(2\pi/L z)$, $J_0=0.2$ and $J_2=0.8$, and c) external input
$I(z)=I_0+\sin(2\pi/L (z-z_0))$, $I_0=1$, $z_0=0.25$.}
\label{fig:ringsetup}
\end{figure}

The stationary solutions obey
\begin{align}
 \partial_\theta \left[1-\cos\theta+(I(z)+ a(z))(1+\cos\theta)\right]\rho(z,\theta)=0\label{eqn:rhostat}\\
 a(z)=2\int w(z,z')\rho(z',\pi) dz'\label{eqn:astat}
\end{align}
The stationary solutions will be qualitatively different depending on the sign of $I+a$.  Consider first the suprathreshold regime where $I+a>0$.  We can then solve (\ref{eqn:rhostat}) to obtain
\begin{align}
 \rho(z,\theta)= \frac{\sqrt{I(z)+a(z)}}{\pi\left[1-\cos\theta+(I(z)+ a(z))(1+\cos\theta)\right]}\label{eqn:rhostatsoln}
\end{align}
which has been normalized such that $\int \rho(z,\theta) d\theta = 1$.  
Inserting this back into (\ref{eqn:astat}) gives
\begin{align}
 a(z)=\frac{1}{\pi}\int w(z,z') \sqrt{I(z)+a(z)} dz'
 \label{eqn:abump}
\end{align}
In the subthreshold regime, $I+a<0$, (\ref{eqn:rhostatsoln}) has a singularity at $1-\cos\theta+(I(z)+ a(z))(1+\cos\theta)=0$, for which there are two solutions $\theta_\pm$ that coalesce in a saddle node bifurcation at $I+a=0$.  Although $\rho$ is no longer differentiable at equilibrium in the subthreshold regime there is still a weak solution. It has been shown previously~\cite{Bard1996} that $\theta_-$ is stable and $\theta_+$ is unstable for a single theta neuron. This implies that the density is given by $\rho(z,\theta)=\delta(\theta-\theta_-)$ and that $\rho(z,\pi)=0$ (i.e. no firing) as expected in the subthreshold regime.  
Figure~\ref{fig:ringsetup}, shows an example of a stationary ``bump" solution for the 
periodic coupling function, $w(z)=-J_0+J_2 \cos(2\pi/L z)$, which has been used
in models of orientation tuning of visual cortex \cite{Sompolinsky1995} and the rodent head direction system~\cite{Zhang2112}).

\section{Beyond mean field theory}
In the infinite $N$ limit when mean field theory applies, the fields $\eta$ and $u$ are completely described by their means.  The time trajectories of these fields are independent of the initial conditions of the individual neurons. For finite $N$, the trajectories can differ for different initial conditions and going beyond mean field theory involves understanding these fluctuations. Implicit in going beyond mean field theory is that these fields are themselves random variables that are drawn from a distribution functional. In this section, we will derive this distribution functional formally and then use it to compute perturbative expressions for the covariances of $\eta$ and $u$.

Recall, that the microscopic system is fully described by
\begin{align}
\partial_t \eta_i(\theta,t) + \partial_\theta F_i(\theta,u_i)\eta_i(\theta,t)-\delta(t-t_0)\eta_i^0(\theta)=0\\
\dot{u}_i(t)+\beta u_i(t)-2\beta \frac{L}{N}\sum_{j=1}^Nw_{ij}\eta_j(\pi,t)-\delta(t-t_0)u_i^0=0
\end{align}
where we have expressed the initial conditions as forcing terms.  The probability density functional for the fields is then comprised of point masses constrained to the dynamical system marginalized over the distribution of the initial data densities:
\begin{align}
{\cal P}[\eta,u]=\int  {\cal D}\eta^0 \,\prod_i \delta\left[\partial_t \eta_i + \partial_\theta F_i(\theta,u_i)\eta_i-\delta(t-t_0)\eta_i^0(\theta)\right]\nonumber\\
\times\delta\left[ \dot{u}_i+\beta u_i-2\beta\frac{L}{N}\sum_jw_{ij}\eta_j(\pi,t)-\delta(t-t_0)u_i^0\right]{\cal P}[\eta^0]
\end{align}
where $P[\eta^0]$ is the probability density functional of the initial neuron densities for all neurons and  ${\cal D}\eta^0$ is the functional integration measure.  We consider the initial condition of $u$ to be fixed to $u_0$.  Using the functional Fourier transform for the Dirac delta functionals, we then obtain
\begin{align}
{\cal P}[\eta,u]=\int {\cal D}\tilde{\eta} {\cal D}\eta^0 {\cal D}\tilde{u}\, &e^{-\sum_i\int dt d\theta\, \tilde{\eta_i}\left[\partial_t \eta_i + \partial_\theta F_i(\theta,u_i)\eta_i-\delta(t-t_0)\eta_i^0(\theta)\right]}\nonumber\\
&\times e^{-\sum_i\int dt \tilde{u_i} \left[ \dot{u}_i+\beta u_i-2\beta\frac{L}{N}\sum_jw_{ij}\eta_j(\pi,t)-\delta(t-t_0)u_i^0\right]}{\cal P}[\eta^0]
\end{align}
where $\tilde\eta_i$ and $\tilde u_i$ are response fields for neuron $i$ with functional integration measures
${\cal D}\tilde{\eta}$ and ${\cal D}\tilde{u}$ over all neurons. If we set $\eta_i^0(\theta)=\delta(\theta-\theta_i(t=0))$, the distribution over initial densities is given by the distribution over the initial phase, $\rho_i^0(\theta)$.  Thus we can write $\int{\cal D}\eta^0 P[\eta^0]=\int \prod_i d\theta \rho_i^0(\theta)$.
The initial condition contribution is given by the integral
\begin{align}
e^{W_0[\tilde{\eta}]}&=\int \prod_i d\theta_i \rho_i^0(\theta_i) e^{\sum_i \tilde{\eta_i}(\theta_i,t_0)}\\
&= \prod_i \int d\theta \rho_i^0(\theta) e^{ \tilde{\eta_i}(\theta,t_0)}\\
&= e^{\sum_i \ln\left( 1- \int d\theta \rho_i^0(\theta) (e^{ \tilde{\eta_i}(\theta,t_0)}-1)\right)}
\end{align}

Hence, the system given by (\ref{eq:drive}) and (\ref{eq:klimontovich}) can be mapped to the distribution functional
${\cal P}[\eta,u]= \int {\cal D}\tilde{\eta}{\cal D}\eta^0 {\cal D}\tilde{u}e^{-S}$ with action $S=S_{\eta}+S_{u}$ given by
\begin{align}
S_{\eta}&=\sum_i \int_{t_0}^{t_1} dt\int_{-\pi}^\pi d\theta \
\tilde{\eta_i}(\theta,t)[\partial_t\eta_i(\theta,t)+ \partial_{\theta}F(\theta,u_i)\eta_i(\theta,t)]+\sum_i\ln\left( 1- \int d\theta\rho_i^0(\theta) (e^{ \tilde{\eta_i}(\theta,t_0)}-1)\right)\\
S_{u}&=\sum_i \int_{t_0}^{t_1} dt \ \tilde{u}_i(t)\left(\dot{u}_i+\beta u_i-2\beta\frac{L}{N}\sum_j w_{ij}\eta_j(\pi,t)-\delta(t-t_0)u_i^0\right)
\end{align}

The exponential in the initial data contribution to the action (which corresponds to a generating function for a Poisson distribution) can be bilinearized via the  Doi-Peliti-Janssen transformation \cite{2007MichaelPRE, MB2010, 2007MichaelPRL, finite2013, Michael2013bymf, MichaelGeneralized2013}: $\psi_i = \eta_i\exp(-\tilde{\eta}_i)$, $\tilde{\psi}_i=\exp(\tilde{\eta_i})-1$, resulting in
\begin{align}
S_{\psi}&=\sum_i\int d\theta dt \tilde{\psi_i}(\theta,t)
[\partial_t\psi_i(\theta,t)+ \partial_{\theta}F(\theta,u_i)\psi_i(\theta,t)]+\sum_i\ln\left( 1- \int d\theta \rho_i^0(\theta)  \tilde{\psi_i}(\theta,t_0)\right)\\
S_{u}&=\sum_i \int dt \tilde{u}_i(t)\left(\dot{u}_i(t)+\beta u_i-\delta(t-t_0)u_i^0-2\beta\frac{L}{N}\sum_j w_{ij}(\tilde{\psi}_j(\pi,t)+1)\psi_j(\pi,t)\right)
\end{align}
where we have not included the noncontributing terms that arise after integration by parts.

We now make the coarse graining transformation $i\rightarrow z N/L$, $u_i(t)\rightarrow u(z,t)$, $\psi_i(\theta,t)\rightarrow \psi(z,\theta,t)$, $\rho_i(\theta,t)\rightarrow \rho(z,\theta,t)$, $I_i\rightarrow I(z)$, $\sum_i \rightarrow (N/L)\int_\Omega dz$, and $w_{ij} \rightarrow w(z-z')$, which yields
\begin{align}
S_{\psi}&=\frac{N}{L}\int dz d\theta dt\, \tilde{\psi}(z,\theta,t)
[\partial_t\psi(z,\theta,t)+ \partial_{\theta}F(\theta,u)\psi(z,\theta,t)]\nonumber\\
&\hspace{50pt}+\frac{N}{L}\int dz\, \ln\left( 1- \int d\theta \rho^0(z,\theta)  \tilde{\psi}(z,\theta,t_0)\right)\\
S_{u}&=\frac{N}{L}\int dz  dt\, \tilde{u}(z,t)\left(\dot{u}+\beta u-\delta(t-t_0)u^0(z)-2\beta\int dz' w(z-z')(\tilde{\psi}(z',\pi,t)+1)\psi(z',\pi,t)\right)
\end{align}

%

\begin{figure}
\includegraphics{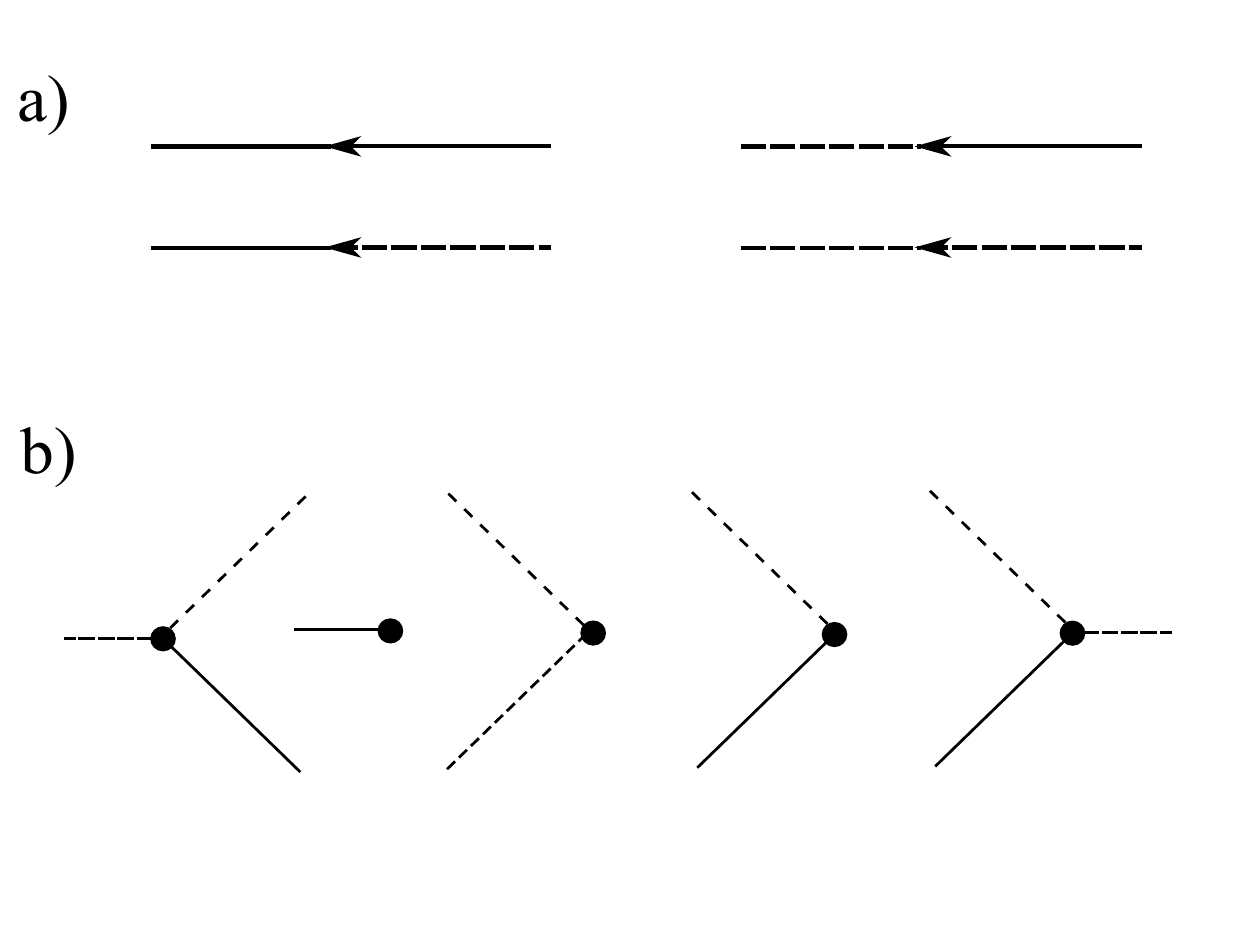}
\caption{a) Propagators, $\Delta_{\nu}^{\nu}(z,t;z',t')$ (upper left), $\Delta_{\nu}^{\varphi}(z,t;z',\theta',t')$ (lower left), $\Delta_{\varphi}^{\nu}(z,\theta,t;z',t')$ (upper right), and $\Delta_{\varphi}^{\varphi}(z,\theta,t;z',\theta', t')$ (lower right) b) Vertices for action in (\ref{action}). From left to right, they are $\partial_{\theta}(1+\cos\theta)$, $0$, $\rho^0(z,\theta)\rho^0(z,\theta)$, $-\frac{2\beta}{N}w(z-z')\rho(z',\pi)$, $-\frac{2\beta}{N}w(z-z')$} 
\label{fig:vertices}
\end{figure}

We examine perturbations around the mean field solutions $a(z,t)$ and $\rho(z,\theta,t)$ of (\ref{eqn:mfcont}) with $u \rightarrow a(z,t)H(t-t_0) +v(z,t)$, $\tilde u \rightarrow \tilde v$, $\psi\rightarrow \rho(z,\theta,t)H(t-t_0) + \varphi(z,\theta,t)$, and $\tilde\psi\rightarrow \tilde\varphi$, where $\rho(z,\theta,t=t_0)=\rho^0(z,\theta)$ and $H(t-t_0)$ is the Heaviside function. We then obtain
\begin{align}
S_{\varphi}&=\frac{N}{L}\int d\theta dtdz\,
\tilde{\varphi} [\partial_t\varphi+ \partial_{\theta}\left[1-\cos\theta+(I+ (a+v))(1+\cos\theta)\right]\varphi+ \partial_{\theta}v(1+\cos\theta)\rho\nonumber\\
&+\frac{N}{L}\int dz \ln\left( 1- \int d\theta \rho^0(z,\theta)  \tilde{\varphi}(z,\theta,t_0)\right)
+\frac{N}{L}\int dz \int d\theta'\tilde{\varphi}(z,\theta',t_0)\rho^0(z,\theta)\nonumber\\
&=\frac{N}{L}\int d\theta dtdz\,
\tilde{\varphi} [\partial_t\varphi+ \partial_{\theta}\left[1-\cos\theta+(I+ (a+v))(1+\cos\theta)\right]\varphi\nonumber\\
&+ \partial_{\theta}v(1+\cos\theta)\rho]-\frac{N}{2L}\int dz \int d\theta'\tilde{\varphi}(z,\theta',t_0)\rho^0(z,\theta)\int d\theta\tilde{\varphi}(z,\theta,t_0)\rho^0(z,\theta)
\end{align}
\begin{align}
S_{v}=\frac{N}{L}\int dtdz\, & \tilde{v}\left[(\frac{d}{dt}+\beta)v-2\beta\int_\Omega dz'\, w(z-z')(\tilde{\varphi}(z',\pi,t)+1)\varphi(z',\pi,t)\right.\nonumber\\
&\left.-2\beta\int_\Omega dz'\, w(z-z')\tilde{\varphi}(z',\pi,t)\rho(z',\pi)-\delta(t-t_0)(u_0(z)-a(z,t_0))\right]
\end{align}
We have only included the quadratic term of the initial condition since it is the only one that plays a role at first order perturbation theory (tree level).
Finally, if we set the mean field solutions to the stationary solutions $\rho(z,\theta)$ and $a(z)$ we obtain
\begin{align}
S_{\varphi}&=\frac{N}{L}\int d\theta dtdz\,
\tilde{\varphi} [\partial_t\varphi+ \partial_{\theta}\left[1-\cos\theta+(I+ (a+v))(1+\cos\theta)\right]\varphi\nonumber\\
&+ \partial_{\theta}v(1+\cos\theta)\rho]+\frac{N}{2L}\int dz \int d\theta'\tilde{\varphi}(z,\theta',t_0)\rho^0(z,\theta)\int d\theta\tilde{\varphi}(z,\theta,t_0)\rho^0(z,\theta)
\end{align}
\begin{align}
S_{v}=\frac{N}{L}\int dtdz\, & \tilde{v}\left[(\frac{d}{dt}+\beta)v-2\beta\int_\Omega dz'\, w(z-z')(\tilde{\varphi}(z',\pi,t)+1)\varphi(z',\pi,t)\right.\nonumber\\
&\left.-2\beta\int_\Omega dz'\, w(z-z')\tilde{\varphi}(z',\pi,t)\rho(z',\pi)\right]
\end{align}
Without loss of generality, we set $L=1$.  In the limit of $N\rightarrow\infty$, the dominant term in the probability density functional for the fields will be the extrema of the action, which defines mean field theory.  Moments of the fields can be computed perturbatively as an expansion in $1/N$ by using Laplace's method around mean field (i.e.~a loop expansion).  The bilinear terms in the action (comprising of a product of a field and a response field) are the linear response functions or propagators.  All the other terms are vertices.  Each vertex contributes a factor of $N$ while each propagator contributes $1/N$.  To make the scaling more transparent, we make the rescaling transformation where $\tilde{v}\rightarrow \tilde{v}/N$ and $\tilde{\varphi}\rightarrow\tilde{\varphi}/N$.  This change will rescale the propagators to order unity and the vertices to order one or higher depending on how many response fields they possess.  The resulting action is 
\begin{align}
S_{\varphi}&=\int d\theta dtdz\,
\tilde{\varphi} [\partial_t\varphi+ \partial_{\theta}\left[1-\cos\theta+(I(z)+ (a(z)+v))(1+\cos\theta)\right]\varphi\nonumber\\
&+ \partial_{\theta}v(1+\cos\theta)\rho]+\frac{1}{2N}\int dz \int d\theta'\tilde{\varphi}(z,\theta',t_0)\rho^0(z,\theta)\int d\theta\tilde{\varphi}(z,\theta,t_0)\rho^0(z,\theta)\nonumber\\
S_{v}&=\int dtdz\,  \tilde{v}\left[\left(\frac{d}{dt}+\beta\right)v-2\beta\int_\Omega dz'\, w(z-z')(\tilde{\varphi}(z',\pi,t)/N+1)\varphi(z',\pi,t)\right.\nonumber\\
&\left.-\frac{2\beta}{N}\int_\Omega dz'\, w(z-z')\tilde{\varphi}(z',\pi,t)\rho(z',\pi)\right]
\label{action}
\end{align}
The propagators and vertices can be represented by Feynman graphs or diagrams (see Fig.~\ref{fig:vertices}).  Each response field corresponds to an outgoing branch (branch on the left) and each field corresponds to an incoming branch (branch on the right).  Time flows from right to left and causality is respected by the propagators.  To each branch is attached a corresponding propagator.

 The propagators are defined by
\begin{eqnarray}
&&G^{-1}\equiv\left(\begin{array}{cc}
\Delta_v^v(x;x') & \Delta_v^\varphi(x;y')\\
\Delta_\varphi^v(y;x') & \Delta_\varphi^\varphi(y;y')
\end{array}\right)^{-1}
=\left.\left(\begin{array}{cc}
\frac{\delta^2S}{\delta \tilde v(x)\delta v(x')}& \frac{\delta^2S}{\delta \tilde v(x)\delta \varphi(y')}\\
\frac{\delta^2S}{\delta \tilde \varphi(y)\delta v(x')}&\frac{\delta^2S}{\delta \tilde \varphi(y)\delta \varphi(y')}
\end{array}\right)\right|_{v,\varphi,\tilde v,\tilde\varphi =0}\\
&=&\left(\begin{array}{cc}
(d/dt +\beta)\delta(x-x') & -2\beta w(z-z')\delta(\pi-\theta')\delta(t-t') \\
\partial_\theta(1+\cos\theta)\rho(z,\theta)\delta(x-x') & (\partial_t+\partial_\theta [1-\cos\theta + (I+ a)(1+\cos\theta)])\delta(y-y')
\end{array}\right)\nonumber\\
\label{eq:propdef}
\end{eqnarray}
where $x=(z,t)$,  and $y=(z,\theta,t)$.  The propagator $\Delta_a^b(x;x')$ is the response of field $a$ at the nonprimed location to field $b$ at the primed location.  The propagator satisfies the condition
\begin{eqnarray}
\int dq'' G^{-1}(q,q'')G(q'',q') =\left(\begin{array}{cc}
\delta(x-x') & 0\\
0 & \delta(y-y')
\end{array}\right)
\label{eq:propcond}
\end{eqnarray}
where $q$ is $x$ or $y$ as appropriate.
Inserting (\ref{eq:propdef}) into (\ref{eq:propcond}) yields
\begin{align}
&(d/dt +\beta)\Delta_v^v(x;x') -2\beta\int dz'' w(z-z'')\Delta_\varphi^v(z'',\pi,t;x')=\delta(x-x')\label{eqn:vv}\\
&(d/dt +\beta)\Delta_v^\varphi(x;y') -2\beta\int dz'' w(z-z'')\Delta_\varphi^\varphi(z'',\pi,t;y')=0\label{eqn:vphi}\\
& (\partial_t+\partial_\theta [1-\cos\theta + (I(z)+ a(z))(1+\cos\theta)])\Delta_\varphi^v(y;x')+\partial_\theta(1+\cos\theta)\rho(z,\theta)\Delta_v^v(x;x') =0\label{eqn:phiv}
\\
& (\partial_t+\partial_\theta [1-\cos\theta + (I(z)+ a(z))(1+\cos\theta)])\Delta_\varphi^\varphi(y;y')+\partial_\theta(1+\cos\theta)\rho(z,\theta)\Delta^\varphi_v(x;y') =\delta(y-y')\label{eqn:phiphi}
\end{align}


\subsection{Computation of Propagators}
In order to perform perturbation theory we must compute the Green's functions or propagators. There are four types of propagators at each spatial location.  The propagator equations are comprised of two sets of $2N$ coupled integro-partial-differential equations.  They can be simplified to ordinary differential equations, which greatly reduces the computational complexity.  The solutions of the equations change qualitatively depending on whether $I+a>0$, suprathreshold regime, and $I+a\le 0$, subthreshold regime. Given that the propagators depend on two coordinates, there are four separate cases.  However, the subthreshold neurons are by definition silent so propagators with the second variable in the subthreshold regime are zero, which leaves two cases for the first variable being supra- or sub- threshold.


\subsubsection{Suprathreshold regime}

In the suprathreshold regime, $z\in \{\zeta : I+a(\zeta)>0\}$, we make the following transformation $\varphi_>: \theta\rightarrow \phi$ where
\begin{equation}
	\phi=\vartheta_>(\theta) = 2 \tan^{-1} \frac{\tan \frac{\theta}{2}}{\sqrt{I(z) + a(z)}}
	\label{eqn:trans}
\end{equation}
which obeys
\begin{equation}
	\frac{d\phi}{d\theta} =\frac{d\vartheta_>(\theta)}{d\theta}= \frac{2 \sqrt{I +  a }}{ (1 - \cos \theta) + (I +  a) ( 1+ \cos \theta) } = 2 \pi \rho(z,\theta)
	\label{eq:jac}
\end{equation}
where the last equality comes from (\ref{eqn:rhostatsoln}).  This transformation has the nice property that $\vartheta_>(\pi)=\pi$.

 Equations (\ref{eqn:vv}) and (\ref{eqn:vphi}) transform to
\begin{align}
&(d/dt +\beta)\Delta_v^v(z,t;z',t') -2\beta\int_> dz'' w(z-z'')\sqrt{I(z'')+a(z'')}\Delta_\varphi^v(z'',\pi,t;z',t')=\delta(z-z')\delta(t-t')\label{eq:lr1}\\
&(d/dt +\beta)\Delta_v^\varphi(z,t;z',\theta',t') -2\beta\int_> dz'' w(z-z'')\sqrt{I(z'')+a(z'')}\Delta_\varphi^\varphi(z'',\pi,t;z',\theta',t')=0\label{eq:lr2}
\end{align}
where we set
$\hat\Delta_\varphi^\cdot(z,\phi,t;\cdot) = \Delta_\varphi^\cdot(z,\vartheta_>^{-1} (\phi),t;\cdot)(d\theta/d\phi)\equiv \Delta_\varphi^\cdot(z,\phi,t,;\cdot)$, where $d\theta/d\phi$ is the Jacobian of the transformation.

Equation (\ref{eqn:phiphi}) transforms to
\begin{align}
(\partial_t&+  \partial_\phi (d\phi/d\theta)[1-\cos\theta + (I+ a)(1+\cos\theta)])\Delta_\varphi^\varphi(z,\phi,t;z',\theta',t')(d\phi/d\theta)\nonumber\\
&+ \partial_\phi (d\phi/d\theta)(1+\cos\theta)\rho(z,\theta)\Delta_v^\varphi(z,t;z',\theta',t') =\delta(z-z')\delta(t-t')\delta(\phi-\vartheta_>(\theta'))(d\phi/d\theta)
\label{eq:fourth}
\end{align}
Now consider
\begin{align}
(1+\cos\theta)\rho(z,\theta)
&=\frac{1}{\pi}\frac{ \sqrt{I +  a(z)}(1+\cos\theta)}{ (1 - \cos \theta) + (I +  a(z)) ( 1+ \cos \theta) } \nonumber\\
&=\frac{1}{\pi}\frac{ \sqrt{I +  a}}{ \tan^2(\theta/2)+ (I +  a)  }  \nonumber\\
&=\frac{1}{\pi}\frac{ \sqrt{I +  a}}{(I +  a)  \tan^2(\phi/2)+ (I +  a)  } \nonumber \\
&=\frac{1}{2\pi}\frac{ 1+ \cos \phi}{\sqrt{I +  a(z)}  }
\label{eq:formula}
\end{align}
where  we have used (\ref{eqn:trans}) and the tangent half-angle formula 
\begin{eqnarray}
\tan^2\frac{\theta}{2}= \frac{1 - \cos \theta}{ 1+ \cos \theta}
\end{eqnarray}
Inserting (\ref{eq:formula}) back into (\ref{eq:fourth}) gives
\begin{align}
(\partial_t+ 2\sqrt{I+ a}) &\partial_\phi \Delta_\varphi^\varphi(z,\phi,t;z',\theta',t')\nonumber\\
&-\frac{\sin\phi}{2\pi\sqrt{I+ a}}\Delta_v^\varphi(z,t;z',\theta',t') =\delta(x-x')\delta(\phi-\vartheta_>(\theta'))\delta(t-t')
\label{eq:lr4}
\end{align}

Similarly, we obtain
\begin{align}
(\partial_t+2\sqrt{I+a}\partial_\phi) \Delta_\varphi^v(z,\phi,t;z',t')-\frac{\sin\phi}{2\pi\sqrt{I+a}}\Delta_v^v(z,t;z',t') =0\label{eq:lr3}
\end{align}

The transformed propagator equations are given by equations (\ref{eq:lr1}), (\ref{eq:lr2}), (\ref{eq:lr4}), and (\ref{eq:lr3}).
Eqs.~(\ref{eq:lr4}) and (\ref{eq:lr3}) are advection equations in $\phi$, which can be integrated to
\begin{align}
\Delta_\varphi^v(z,\phi,t;z',t')=C(z)\int_{t'}^td\tau&\sin(\phi-\nu_>(z)(t-\tau))\Delta_v^v(z,\tau;z',t')\nonumber\\
\Delta_\varphi^\varphi(z,\phi,t;z',\theta',t')=C(z)\int_{t'}^td\tau&\sin(\phi-\nu_>(z)(t-\tau))\Delta_v^\varphi(z,\tau;z',\theta',t')\nonumber\\
&+\delta(\phi-\vartheta_>(\theta')-\nu_>(z)(t-t'))\delta(z-z')
\end{align}
where
\begin{align}
C(z)&\equiv\frac{1}{2\pi\sqrt{I(z)+a(z)}}\nonumber\\
\nu_>(z)&\equiv2\sqrt{I(z)+a(z)}
\end{align}
We then define the following variables:
\begin{align}
r^v(z,t;z',t')&=\Delta_\varphi^v(z,\pi,t;z',t')=C(z)\int_{t'}^td\tau\sin(\nu_>(z)(t-\tau))\Delta_v^v(z,\tau;z',t')\\
r^{\varphi}(z,t;z',\theta',t')&=\Delta_\varphi^\varphi(z,\pi,t;z',\theta',t')-\delta(\pi-\vartheta_>(\theta')-\nu_>(z)(t-t'))\delta(z-z')\nonumber\\
&=C(z)\int_{t'}^td\tau\sin(\nu_>(z)(t-\tau))\Delta_v^\varphi(z,\tau;z',\theta',t')
\end{align}
We thus obtain after repeated derivatives and using the propagator equations  (\ref{eq:lr1}), (\ref{eq:lr2}), (\ref{eq:lr4}), and (\ref{eq:lr3}):
\begin{align}
&\frac{d^2}{dt^2}r^v(z,t;z',t')=\frac{1}{\pi}\Delta_v^v(z,t;z',t')-\nu_>^2(z)r^v(z,t;z',t')\label{eq:lrODE1}\\
&(\frac{d}{dt}+\beta)\Delta_v^v(z,t;z',t')-\beta \int dz''w(z-z'')\nu_>(z'')r^v(z'',t'';z',t')=\delta(z-z')\delta(t-t^{\prime})\\
&\frac{d^2}{dt^2}r^{\varphi}(z,t;z',\theta',t')=\frac{1}{\pi}\Delta_v^{\varphi}(z,t;z',\theta',t')-\nu_>^2(z)r^{\varphi}(z,t;z',\theta',t')\label{eq:lrODE3}\\
&(\frac{d}{dt}+\beta)\Delta_v^{\varphi}(z,t;z',\theta',t')-\beta\int dz''w(z-z'')\nu_>(z'')r^{\varphi}(z'',t'';z',\theta',t')\nonumber\\
&=\beta w(z-z')\nu_>(z')\delta(\pi-\vartheta_>(\theta')-\nu_>(z')(t-t'))\label{eq:lrODE4}
\end{align}

The covariance function (\ref{eqn:cov}) involves the integral quantity  
\begin{align}
U(z,t;z',t_0)&=\int d\theta'\Delta_v^{\varphi}(z,t;z',t_0,\theta')\rho^0(z',\theta')
\end{align}
by our choice of transformation convention.
However, instead of computing the propagator at all values of $\theta'$, we create another pair of ODEs for $U$.  Applying the integral operator $\int d\theta'\rho^0(z',\theta')$ to (\ref{eq:lrODE3}) and (\ref{eq:lrODE4}) gives
\begin{align}
\frac{d^2}{dt^2}r(z,t;z',t')&=\frac{1}{\pi}U(z,t;z',t')-\nu_>^2(z)r(z,t;z',t')\\
(\frac{d}{dt}+\beta)U(z,t;z',,t')&-\beta\int_> dz''w(z-z'')\nu_>(z'')r(z'',t'';z',t')\nonumber\\
&=\beta w(z-z')\nu_>(z')\int d\theta'\rho^0(z',\theta')\delta(\pi-\vartheta(\theta')-\nu_>(z')(t-t'))\nonumber\\
&=\beta w(z-z')\nu_>(z')\rho^0(z',\theta_0)\left.\frac{d\theta}{d\phi}\right|_{\theta'=\theta_0}\nonumber\\
&=\frac{\beta}{2\pi}w(z-z')\nu_>(z')\frac{\rho^0(z',\theta_0)}{\rho(z',\theta_0)}
\end{align}
where $r(z,t;z',t')=\int r^{\varphi}(z,t;z',\theta',t')\rho^0(z',\theta')\, d\theta'$ and $\theta_0=-\vartheta^{-1}(\nu_>(z')(t-t')))$.
Hence, we need to numerically integrate the following equations
\begin{align}
&\frac{d^2}{dt^2}r(z,t;z',t')=\frac{1}{\pi}U(z,t;z',t')-\nu_>^2(z)r(z,t;z',t')\label{eqn:begin}\\
&(\frac{d}{dt}+\beta)U(z,t;z',,t')-\beta\int_> dz''w(z-z'')\nu_>(z'')r(z'',t'';z',t')=\frac{1}{2\pi}\beta w(z-z')\nu_>(z')\\
&\frac{d^2}{dt^2}r^v(z,t;z',t')=\frac{1}{\pi}\Delta_v^v(z,t;z',t')-\nu_>^2(z)r^v(z,t;z',t')\\
&(\frac{d}{dt}+\beta)\Delta_v^v(z,t;z',t')-\beta \int_> dz''w(z-z'')\nu_>(z'')r^v(z'',t'';z',t')=\delta(z-z')\delta(t-t^{\prime})\\
&\frac{d^2}{dt^2}r^{\varphi}(z,t;z',\pi,t')=\frac{1}{\pi}\Delta_v^{\varphi}(z,t;z',\pi,t')-\nu_>^2(z)r^{\varphi}(z,t;z',\pi,t')\\
&(\frac{d}{dt}+\beta)\Delta_v^{\varphi}(z,t;z',\pi,t')-\beta\int_> dz''w(z-z'')\nu_>(z'')r^{\varphi}(z'',t'';z',\pi,t')\nonumber\\
&=\frac{\beta}{2} w(z-z')\delta(t-t')+\beta\sum_{l=1}^\infty w(z-z')\delta(t-t^{\prime}-T_l(z'))\label{eqn:last}
\end{align}
where $T_l(z')=\{s | \nu(z) s=2\pi\}$ marks the time intervals from $t'$ such that $2\pi l-\nu_>(z')T_l(z')=0$.  The source at $t=t'$ in (\ref{eqn:last}) has a factor of one half because because it comes form the $\theta$ delta function, which is symmetric about $\theta=\theta'$ since the propagator is symmetric at $\theta=\theta'$, unlike the contribution from the time delta function, which is one sided due to causality.

\subsubsection{Subthreshold regime}
In the subthreshold regime, namely $I+a\leq 0$, the mean field solution for the density $\rho$ is a point mass, and this will change the form of the propagators.  The propagator equations are
\begin{align}
&(d/dt +\beta)\Delta_v^v(x;x') -2\beta\int dz'' w(z-z'')\Delta_\varphi^v(z'',\pi,t;x')=\delta(x-x')\\
&(d/dt +\beta)\Delta_v^\varphi(x;y') -2\beta\int dz'' w(z-z'')\Delta_\varphi^\varphi(z'',\pi,t;y')=0\\
& (\partial_t+\partial_\theta [1-\cos\theta + (I(z)+ a(z))(1+\cos\theta)])\Delta_\varphi^v(y;x')\nonumber\\
&+\partial_\theta(1+\cos\theta)\delta(\theta-\theta_-(z))\Delta_v^v(x;x') =0\\
& (\partial_t+\partial_\theta [1-\cos\theta + (I(z)+ a(z))(1+\cos\theta)])\Delta_\varphi^\varphi(y;y')\nonumber\\
&+\partial_\theta(1+\cos\theta)\delta(\theta-\theta_-(z))\Delta^\varphi_v(x;y') =\delta(y-y')
\end{align}
where the equations are defined on $I(z)+a(z)<0$, and $\theta_\pm$ are the mean field fixed points, where $\sin\theta_\pm=\pm 2\sqrt{|I+a|}/(1+|I+a|)$.  However, note that the primed variables are defined over the entire $z$ domain since subthreshold neurons can receive input from suprathreshold neurons.

We simplify these equations by breaking the domain of $\theta$ into two pieces: $D1=(\theta_+,\theta_-)$ and $D2=(\theta_-,\theta_+)$.  In the two advection equations, there will be a clockwise advection of the propagators towards $\theta_-$ in D1 and in a counterclockwise advection towards $\theta_-$ in D2.  $\pi$ is in D1 but not D2 so neurons starting in D2 will never fire.  In D1, we make the transformation $\vartheta_<: \theta\rightarrow\chi$:
\begin{align}
\chi=\vartheta_<(\theta)=\ln\left(\frac{\sin\theta-\sqrt{|I+ a|}(1+\cos\theta)}{\sin\theta+\sqrt{|I+ a|}(1+\cos\theta)}\right)
\end{align}
\begin{align}
\chi=2\coth^{-1}\frac{\tan\frac{\theta}{2}}{ \sqrt{|I(z)+a(z)|}}
\end{align}
\begin{align}
 \frac{d\chi}{d\theta} = \frac{2 \sqrt{|I +  a|}}{ (1 - \cos \theta) - |I +  a| ( 1+ \cos \theta) } 
\end{align}
\begin{align}
 \frac{d\theta}{d\chi} = \frac{\sqrt{|I +  a|}}{ (|I +  a|^2+1)\cosh^2(\chi/2)-1 } 
\end{align}
\begin{align}
 \frac{d\theta}{d\chi} = \frac{2\sqrt{|I +  a|}}{ (|I +  a|^2+1)(\cosh(\chi)+1)-2 } 
\end{align}
\begin{align}
1+\cos\theta = \frac{2}{1+|I+a|\coth^2\chi/2}
\end{align}
which maps D1 to the real line where $-\infty$ corresponds to $\theta_+$ and $\infty$ corresponds to $\theta_-$.  

We then have the following propagator equations in the $\chi$ representation:
\begin{align}
&(d/dt +\beta)\Delta_v^v(z,t;z',t')-\beta\int_> dz'' w(z-z'')\nu_>(z'')\Delta_\varphi^v(z'',\pi,t;z',t')=\delta(z-z')\delta(t-t')\\
&(\partial_t+\nu_<\partial_\chi )\Delta_\varphi^v(z,\chi,t;z',t')=Q(z,\chi)\Delta_v^v(z,t;z',t')\\
&(d/dt +\beta)\Delta_v^\varphi(z,t;z',\theta',t') -\beta\int_> dz'' w(z-z'')\nu_>(z'')\Delta_\varphi^\varphi(z'',\pi,t;z,\theta',t')=0\\
&(\partial_t+\nu_<\partial_\chi )\Delta_\varphi^\varphi(z,\chi,t;z',\theta',t')
=Q(z,\chi)\Delta_v^\varphi(z,t;z',\theta',t')+\delta(z-z')\delta(\chi-\vartheta_<(\theta'))\delta(t-t')
\end{align}
where  
$$Q(z,\chi)=-\partial_\chi \frac{2}{1+|I+a|\coth^2\chi/2}\delta\left(\vartheta^{-1}(\chi)-\theta_-(z)\right)$$
 and $\nu_<(z)=2\sqrt{|I(z)+a(z)|}$.
Integrating yields
\begin{align}
\Delta_\varphi^v(z,\chi,t;w')&=\int_{t'}^t Q(z,\chi-\nu_<(z)(t-\tau))\Delta_v^v(z,\tau;z',t') d\tau\\
\Delta_\varphi^\varphi(z,\chi,t;w')&=\int_{t'}^t Q(z,\chi-\nu_<(z)(t-\tau))\Delta_v^\varphi(z,\tau;z',\theta',t')  d\tau\nonumber\\
&+ \delta(z-z')\delta(\chi-\vartheta_s(\theta')-\nu_<(z)(t-t'))
\end{align}
Hence, the only contribution from the subthreshold neurons are from  any neuron that is initially in D1, which for uniformly distributed phases the probability will be $(1 -(\theta_+-\theta_-)/2\pi$.  The subthreshold propagators are thus passively driven by the superthreshold propagators.  Hence, for $z$ in the subthreshold regime, the relevant propagator equations are:
\begin{align}
(d/dt +\beta)\Delta_v^v(z,t;z',t')& -\beta\int_> dz'' w(z-z'')\nu_>(z'')r^v(z'',t;z',t')\\
&=\delta(z-z')\delta(t-t')
\end{align}
\begin{align}
(d/dt +\beta)\Delta_v^\varphi(z,t;z',\pi,t')& -\beta\int_> dz'' w(z-z'')\nu_>(z'')r^\varphi(z'',t;z',\pi,t')\nonumber\\
&=\frac{\beta}{2}w(z-z')\delta(t-t^{\prime})+\beta\sum_{l=1}^\infty w(z-z')\delta(t-t^{\prime}-T_l(z'))
\end{align}
\begin{align}
(d/dt +\beta)U(z,t;z',t')& -\beta\int_> dz'' w(z-z'')\nu_>(z'')r(z'',t;z',t')\nonumber\\
&=\frac{\beta}{2\pi}w(z-z')\nu_>(z')
\end{align}

\subsection{Covariance functions}
\subsubsection{Drive covariance}

\begin{figure}
\includegraphics{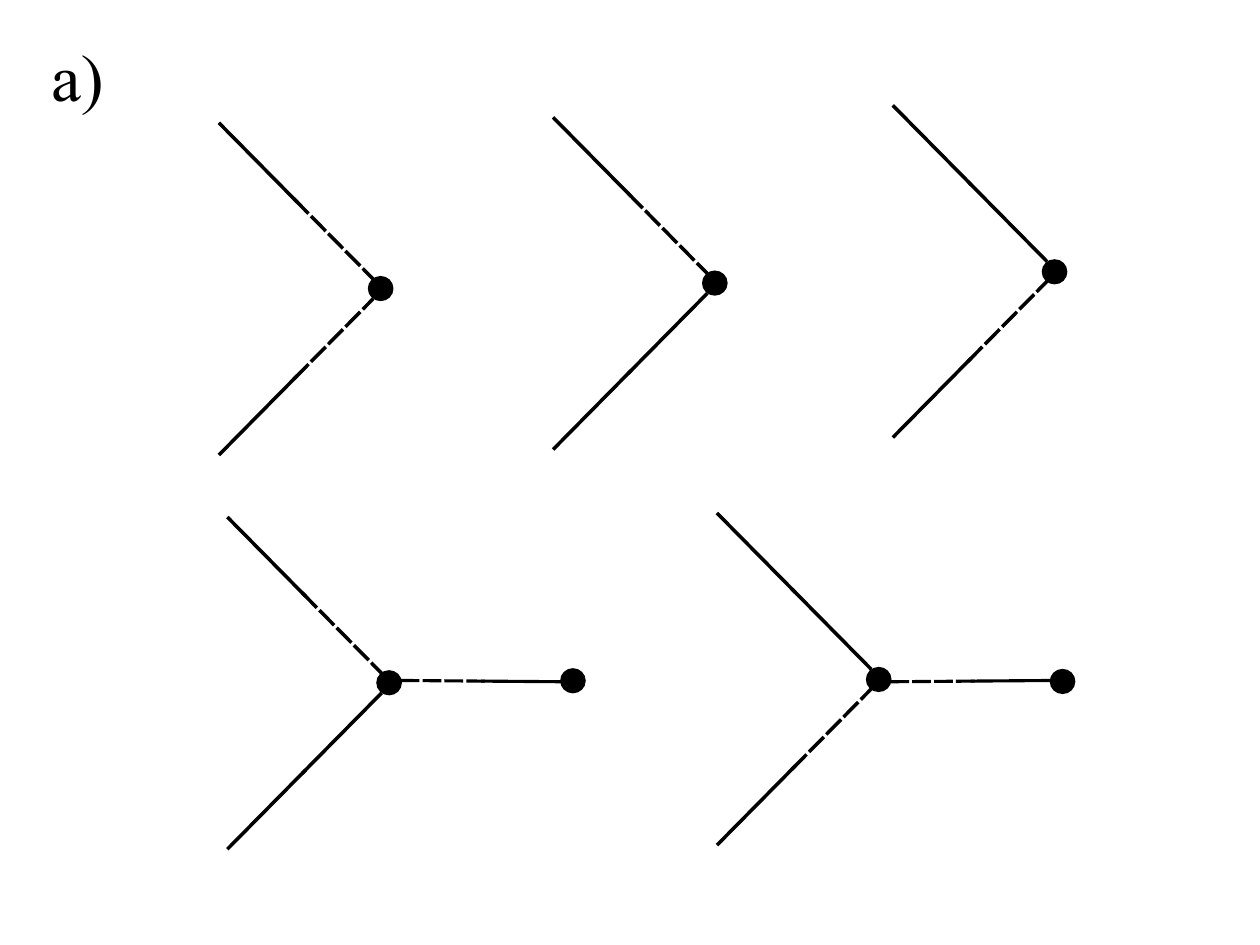}
\includegraphics{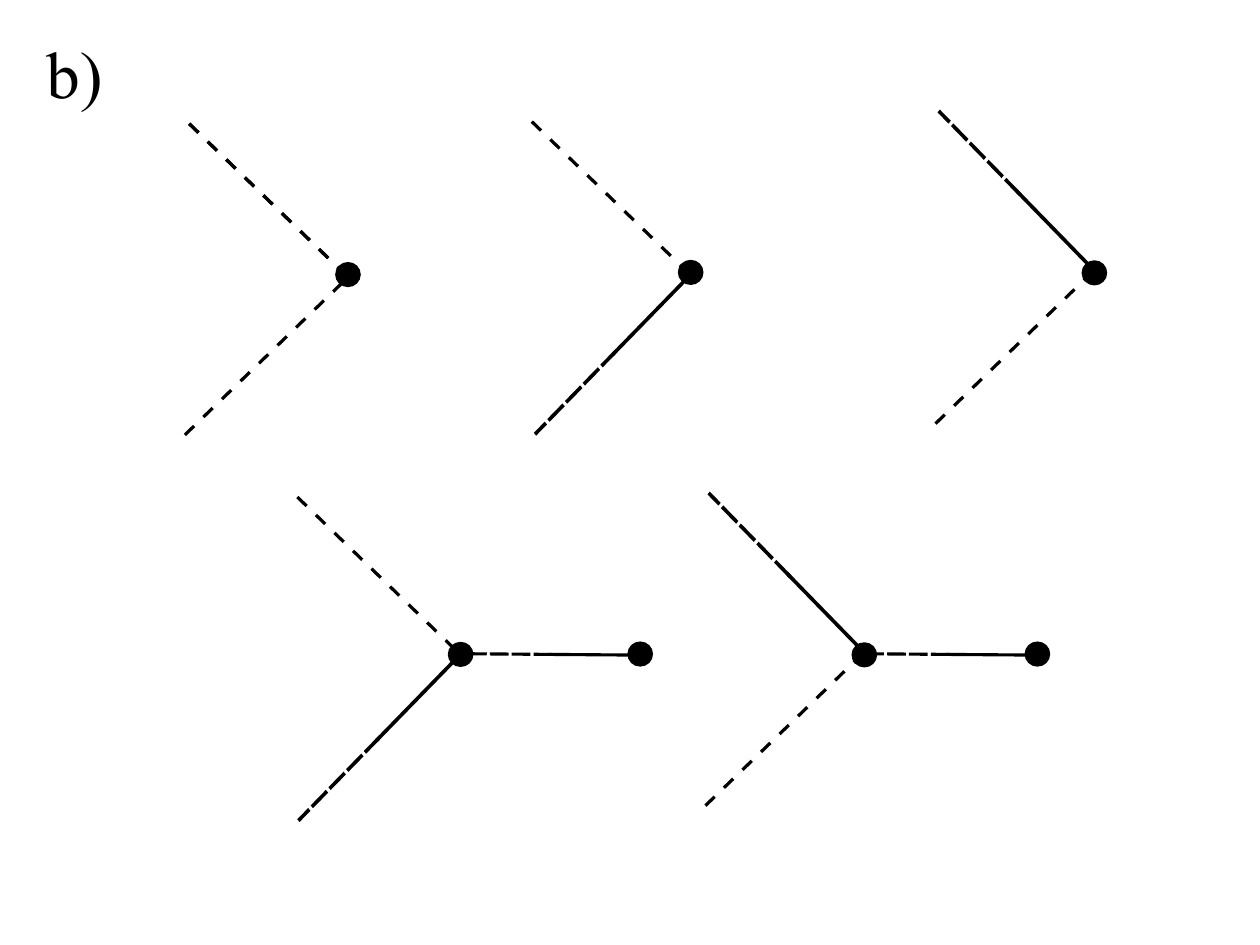}
\caption{Tree level diagrams for a) drive covariance $\langle v v \rangle$ and b) rate covariance $ \langle \varphi \varphi \rangle$. The lower two diagrams are zero for a) and b). For the upper 3 diagrams in a) and b), the first diagram corresponds to the third term, while the second and third diagrams correspond to the first and second terms of eq.~\ref{eqn:cov} and eq.~\ref{eqn:ratecov} respectively.} 
\label{fig:diagrams}
\end{figure}

As described previously~\cite{finite2013}, the covariances between the fields to order $1/N$ are comprised of vertices with two outgoing branches.  Using the diagrams in Figures~\ref{fig:vertices} and \ref{fig:diagrams}, we obtain
%
%
%
%
%
%
%
%
\begin{align}
N\langle  \delta v(z,t) \delta v(z',t')\rangle&=2\beta\int dz_1dz_2 d\tau \Delta_v^v(z,t;z_1,\tau)\Delta_v^{\varphi}(z',t';z_2,\pi,\tau)w(z_1-z_2)\rho(z_2,\pi)\nonumber\\
&+(x\leftrightarrow x')-\int dz_1\left\{\int d\theta\Delta_v^{\varphi}(z,t;z_1,t_0,\theta)\rho(z_1,\theta,t_0)\right.\nonumber\\
&\times\left.\int d\theta'\Delta_v^{\varphi}(z',t';z_1,t_0,\theta')\rho(z_1,\theta', t_0)\right\}
\label{eqn:cov}
\end{align}

Evaluating the covariance function in (\ref{eqn:cov}) requires computing the propagators using the equations derived in the previous section. Our numerical methods for integrating these equations are in the Appendix.
We compared the theory to microscopic simulations of~(\ref{eq:drive1}) with fixed initial condition of $u(z)$ set to the mean field solution $a(z)$,  and the initial condition of $\theta(z)$ is sampled from the probability distribution obeying the mean field solution $\rho(z,\theta)$.
For the supra-threshold region, the cumulative distribution function for $\rho(z,\theta)$ is
\begin{align}
P(z,\theta)= \frac{1}{\pi}\tan^{-1}\left(\frac{\sin\theta}{\sqrt{I(z)+ a(z)}(1+\cos\theta)}\right)+\frac{1}{2}
\label{cdf}
\end{align} 
from which we can sample $\theta$ by applying the inverse of (\ref{cdf}) to a uniform random number.
For the sub-threshold region, all the samples are taken to be at the stable solution $\theta_-(z)=-2\tan^{-1}(\sqrt{(|I(z)+a(z)|)})$. 

\begin{figure}[t!]
\begin{minipage}{.52\textwidth}
\includegraphics[width=1\linewidth, height=0.4\textheight]{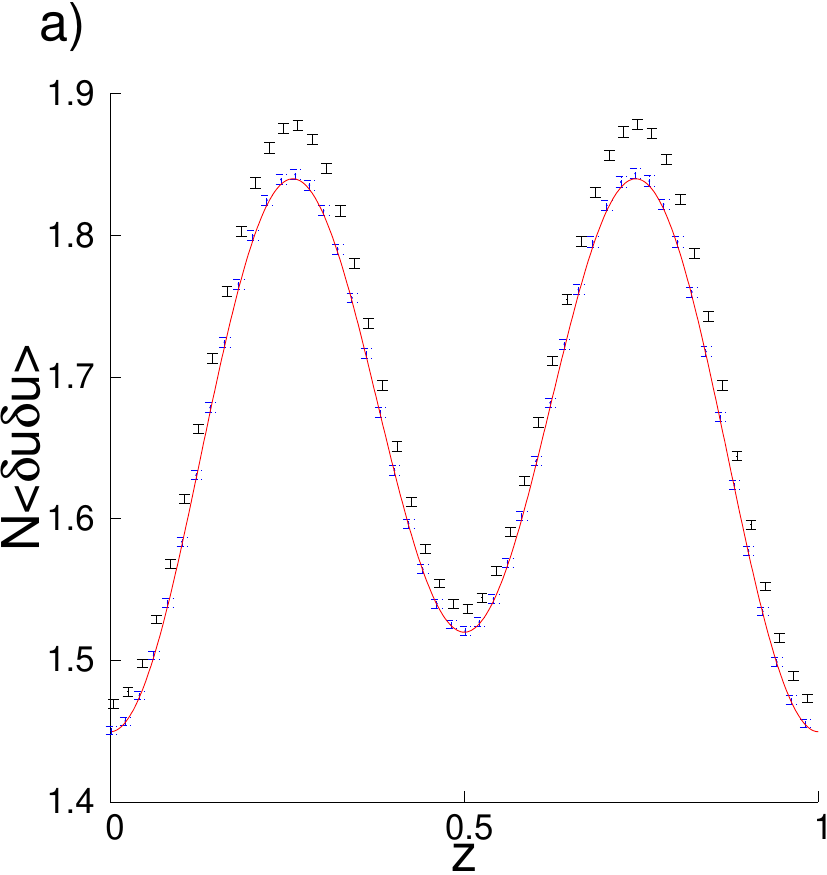}
\end{minipage}
\begin{minipage}{0.9\textwidth}
\includegraphics[width=1\linewidth, height=0.2\textheight]{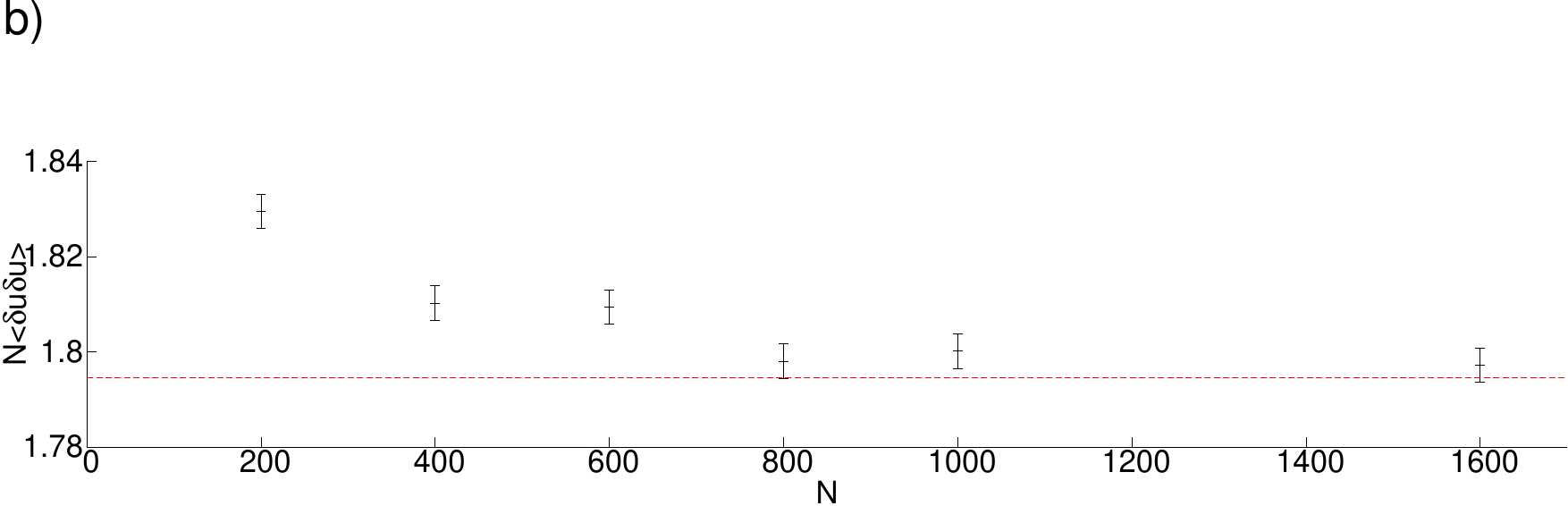}
\end{minipage}
\caption{ Variance times $N$ at time $t=10$ for parameters in figure~\ref{fig:ringsetup}.   
a) Comparison between microscopic simulation and theory calculation for $N=200$ and $N=800$.
b) The $N$ dependence of $N\langle \delta u(z)u(z)\rangle$ at $z=0.2$. Standard errors for the microscopic simulation are estimated by bootstrap.}
\label{fig:ringscale}
\end{figure}

\begin{figure}[ht]
\begin{minipage}{.48\textwidth}
\includegraphics[width=1\linewidth, height=0.3\textheight]{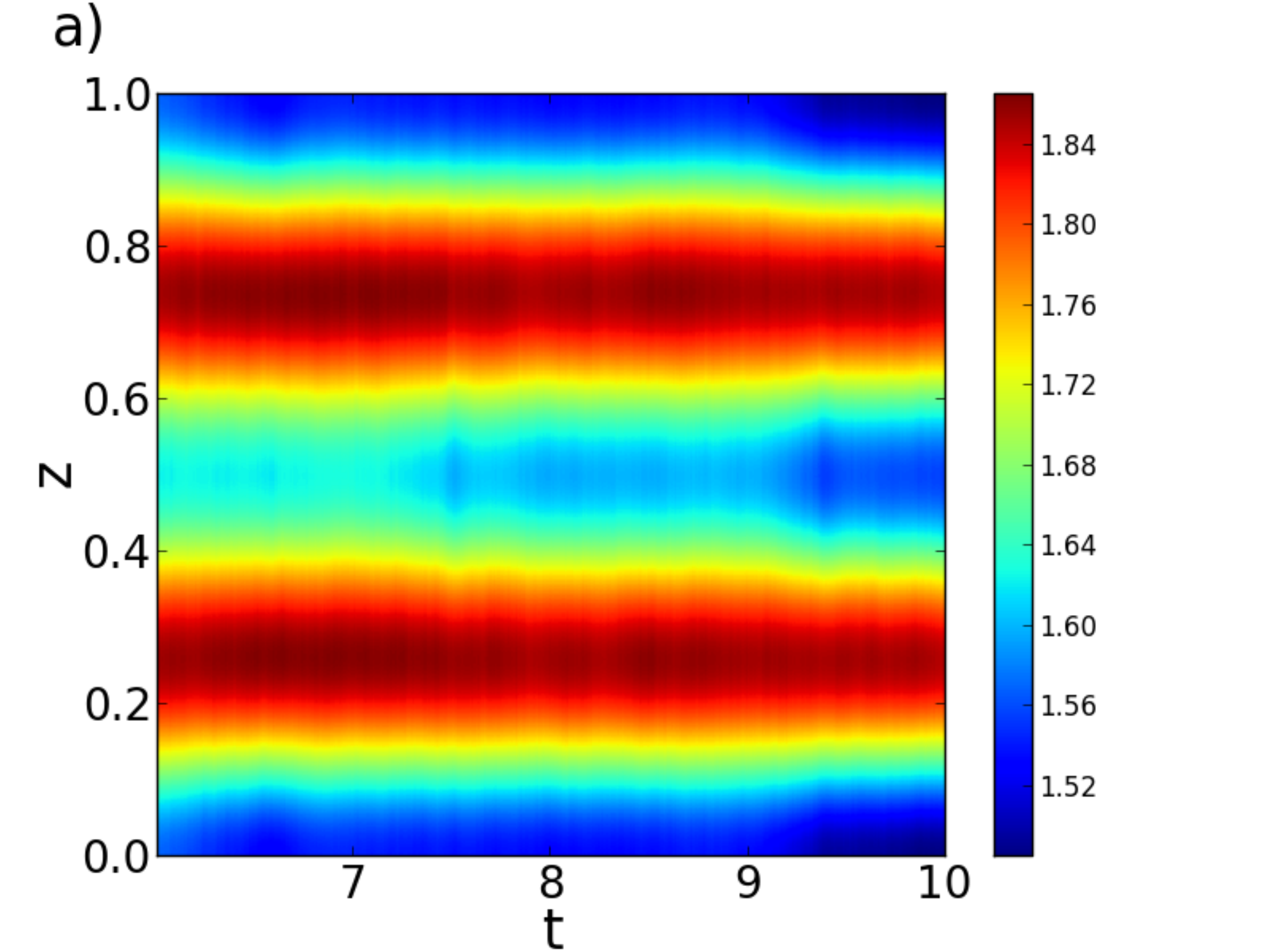}
\end{minipage}
\begin{minipage}{.48\textwidth}
\includegraphics[width=1\linewidth, height=0.3\textheight]{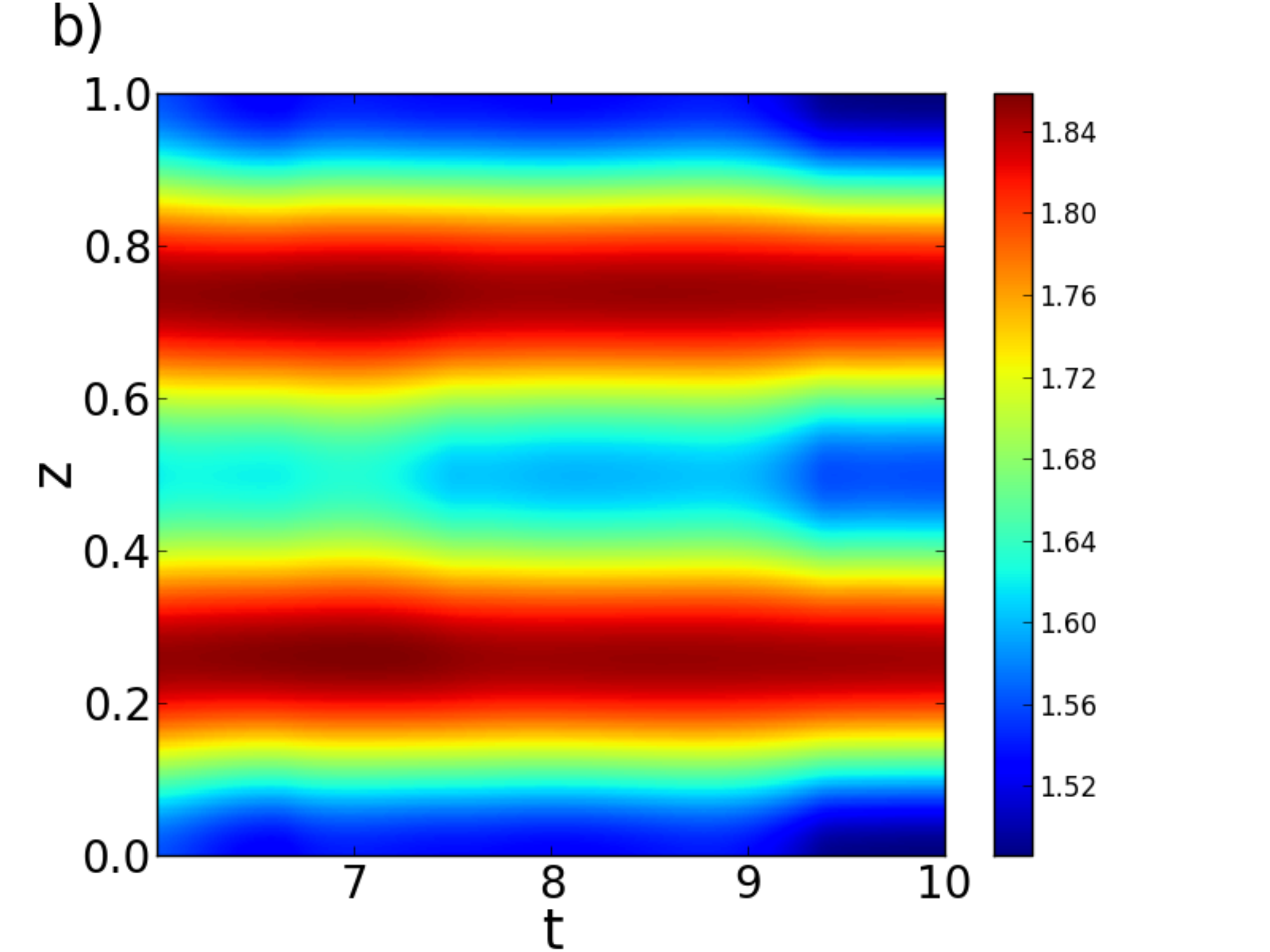}
\end{minipage}
\begin{minipage}{.48\textwidth}
\includegraphics[width=1\linewidth, height=0.3\textheight]{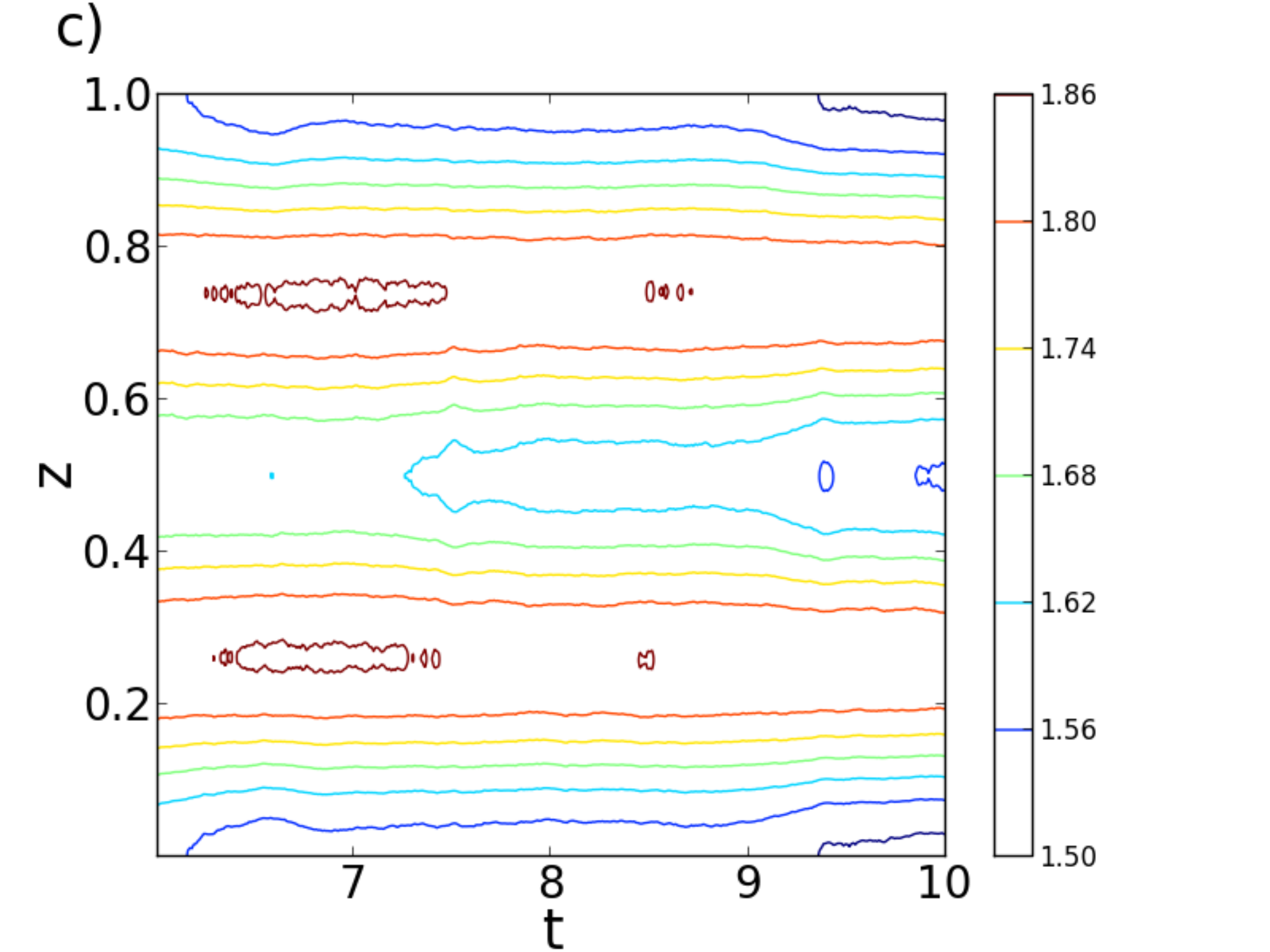}
\end{minipage}
\begin{minipage}{.48\textwidth}
\includegraphics[width=1\linewidth, height=0.3\textheight]{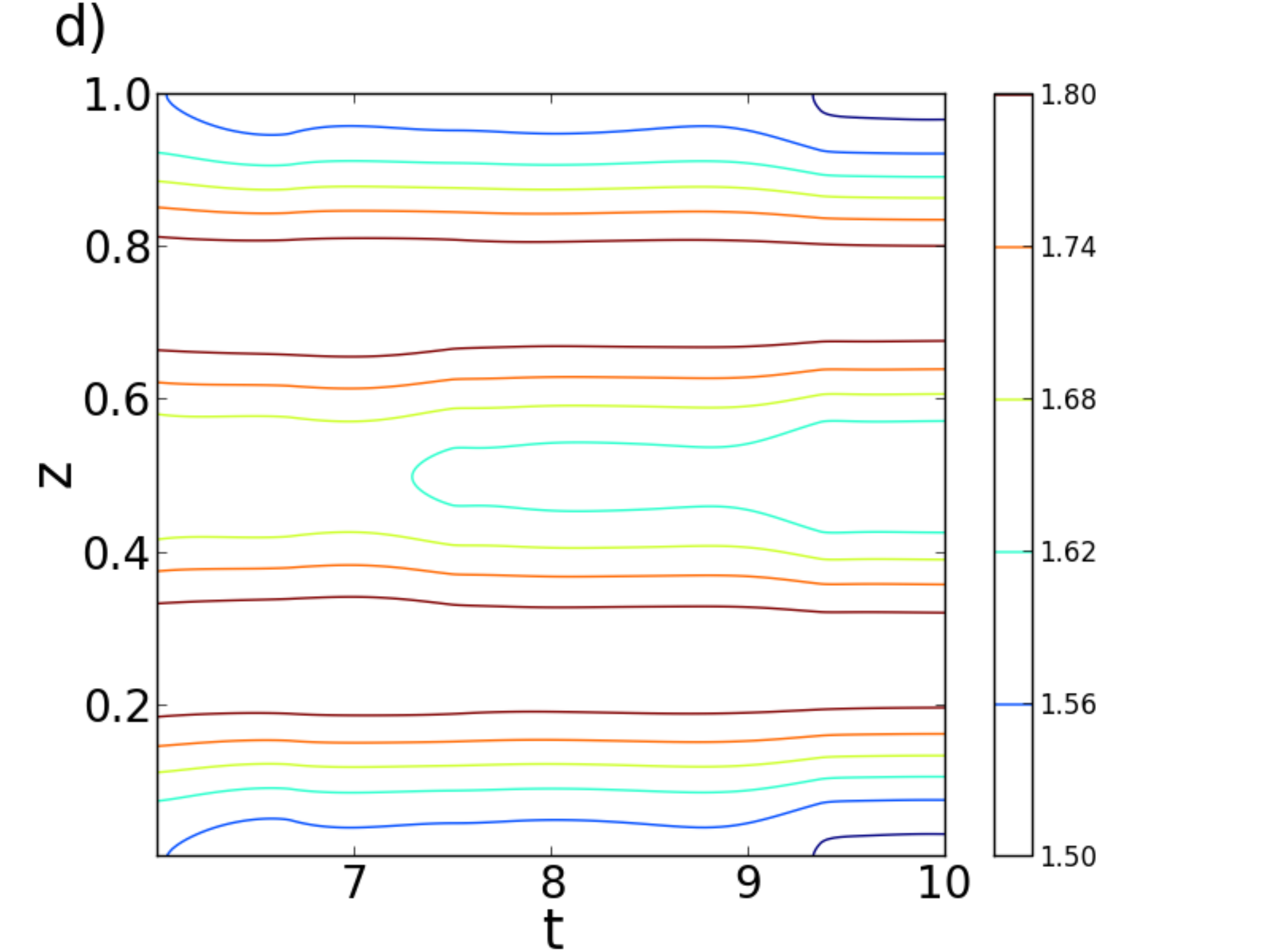}
\end{minipage}
\caption {Spatial-temporal dynamics of the synaptic drive variance for the microscopic simulation for $N=800$ in a) and c),  and the theory in b) and d). Parameters are as in Figure 1.}
\label{fig:HeatMap}
\end{figure}

\begin{figure}[t!]
\begin{minipage}{0.45\textwidth}
\includegraphics[width=1\linewidth, height=0.2\textheight]{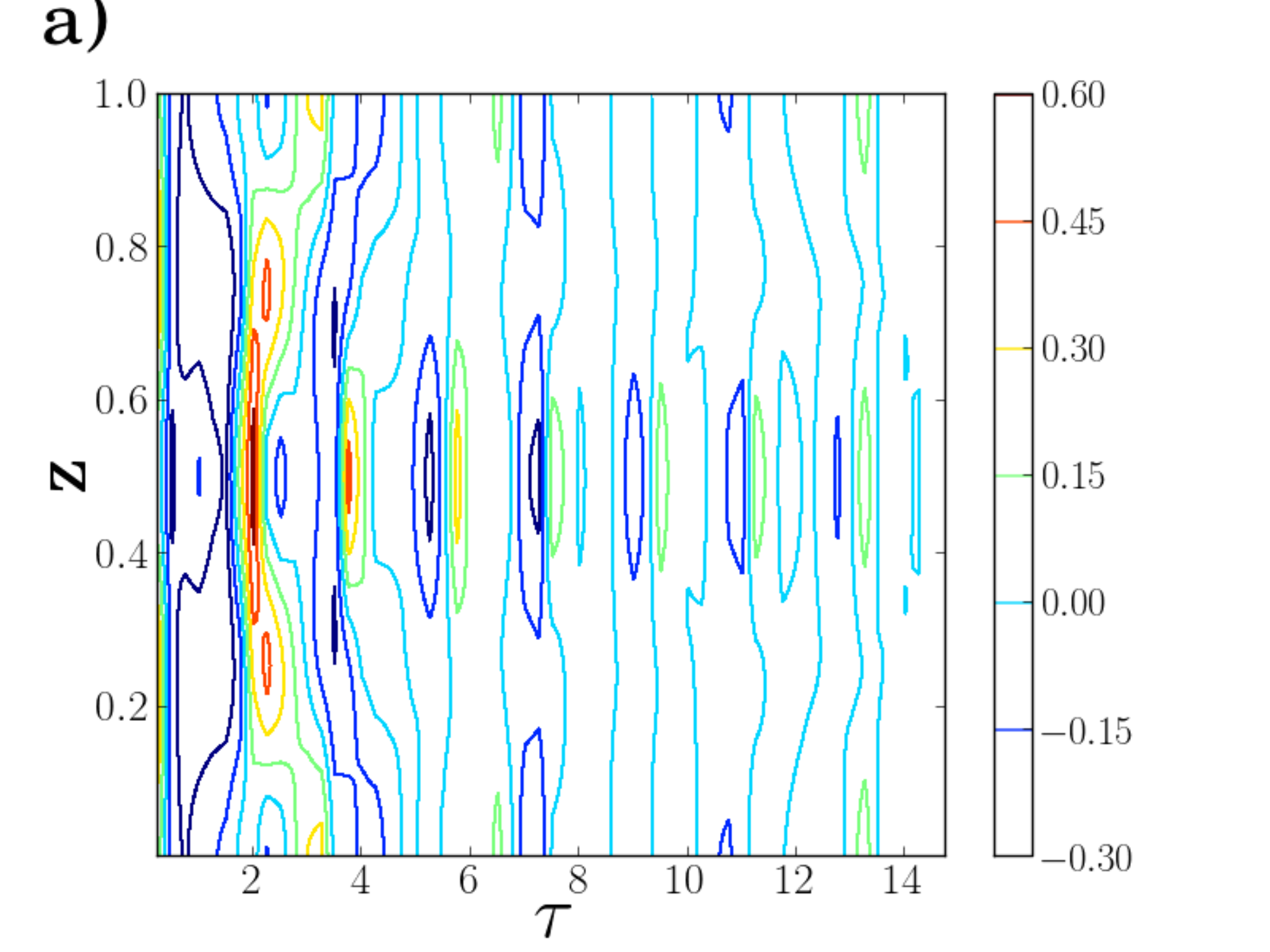}
\end{minipage}
\begin{minipage}{.45\textwidth}
\includegraphics[width=1\linewidth, height=0.2\textheight]{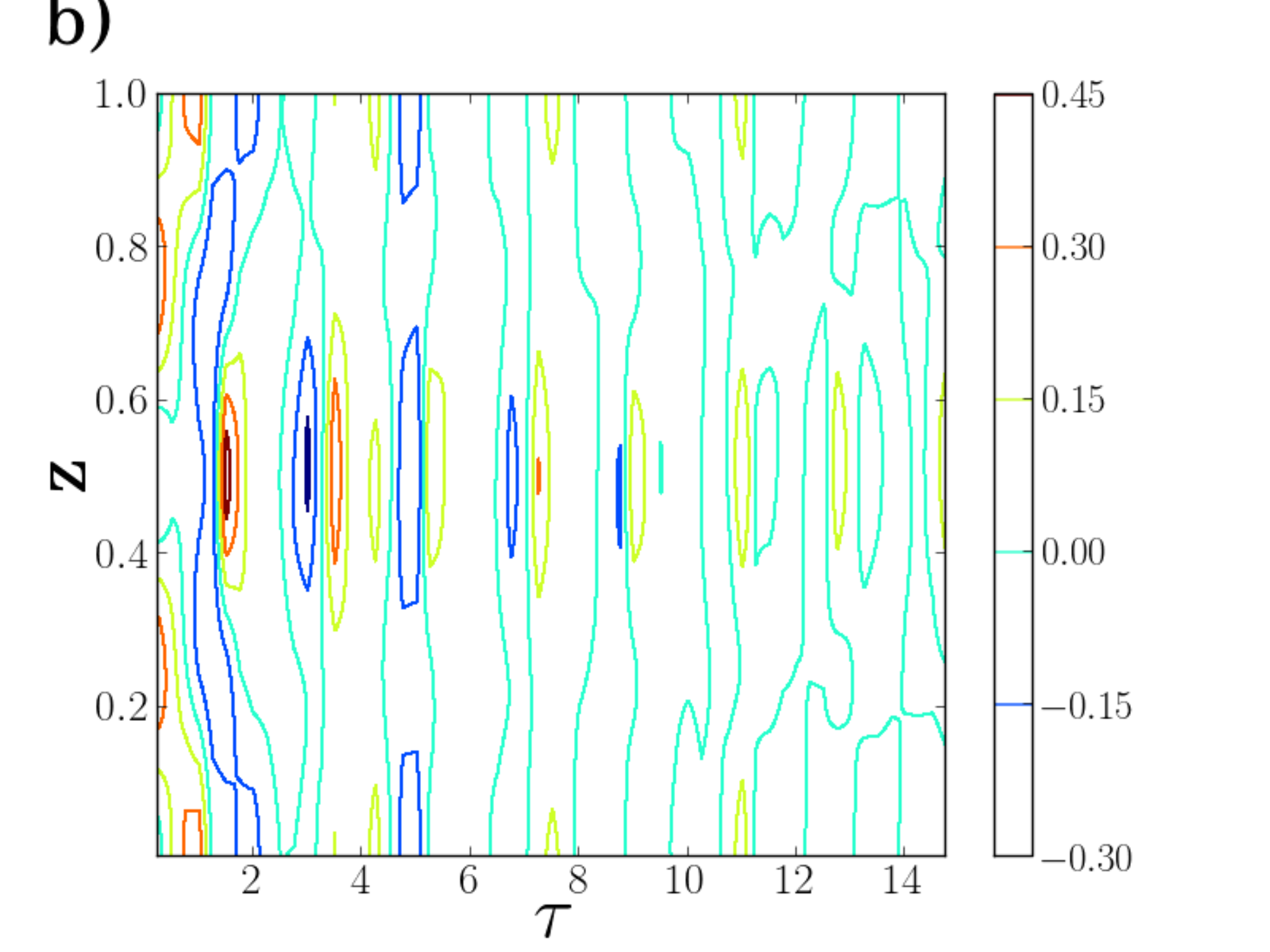}
\end{minipage}
\begin{minipage}{.45\textwidth}
\includegraphics[width=1\linewidth, height=0.2\textheight]{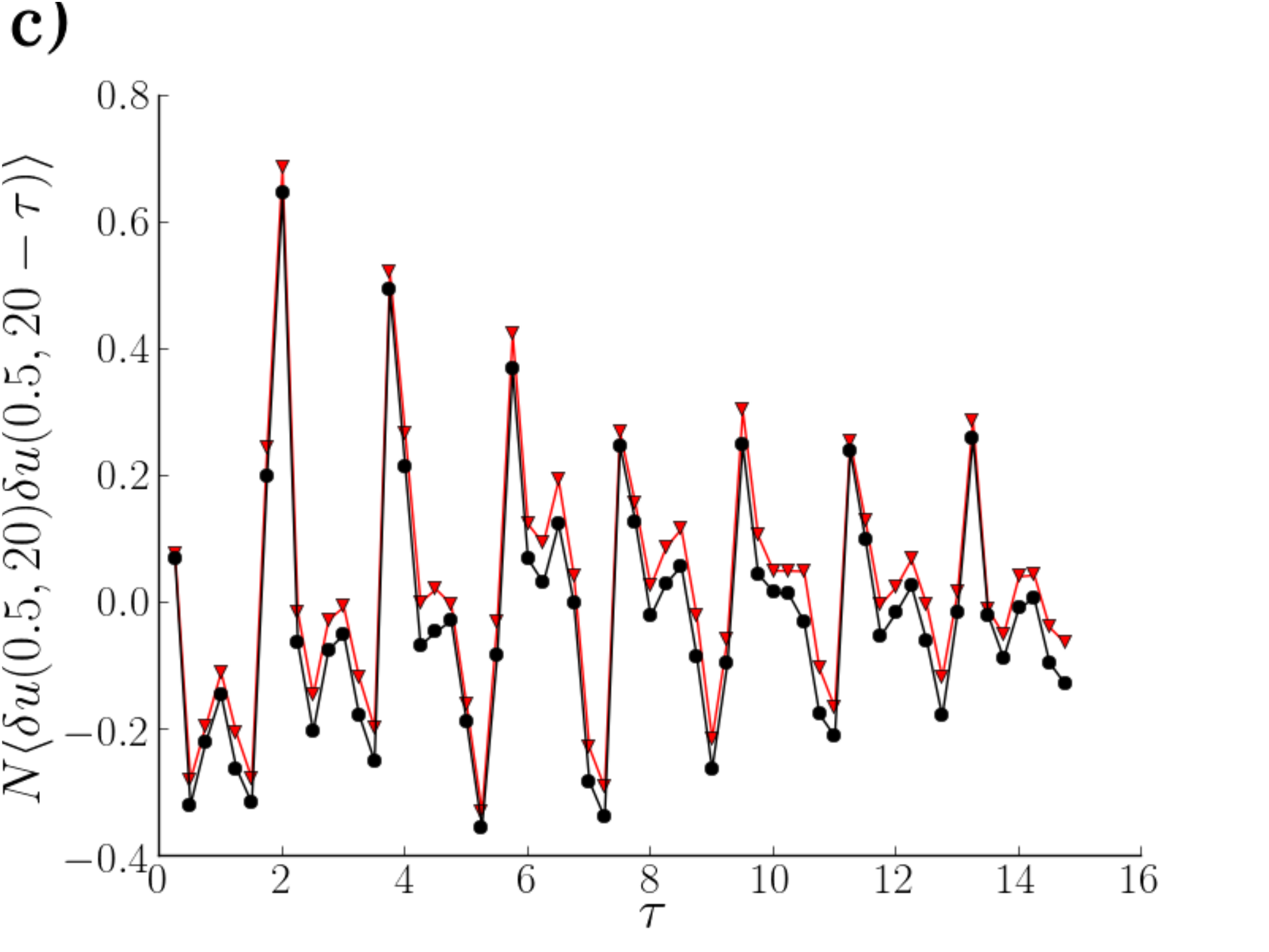}
\end{minipage}
\begin{minipage}{.45\textwidth}
\includegraphics[width=1\linewidth, height=0.2\textheight]{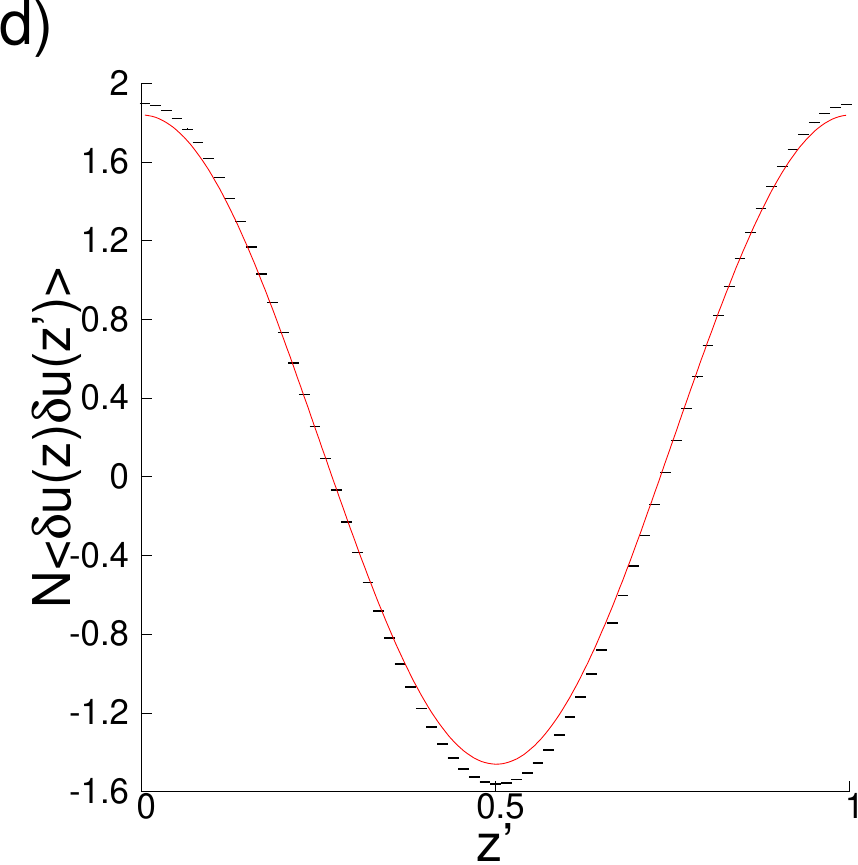}
\end{minipage}
\caption { Spatiotemporal plot of covariance $\langle\delta u(z,20)\delta u(z,20-\tau)\rangle$ for a) theory and b) microscopic simulation using parameters from Figure~\ref{fig:ringsetup}. c) Covariance at a single spatial location, $\langle\delta u(0.5, 20)\delta u(0.5, 20-\tau)\rangle$. d) Covariance at a single time, $\langle\delta u(z=.005, 20)\delta u(z',20)\rangle$. Standard errors are estimated by jackknife.}
\label{fig:multidimension}
\end{figure}

A comparison between the variance of synaptic drive fluctuations for the microscopic simulation as a function of space at a fixed time for two values of $N$ and the theory  is shown in figure~\ref{fig:ringscale}a) for external input and synaptic coupling weight as in figure~\ref{fig:ringsetup}.  This is a case where all neurons are in the supra-threshold region.   We see that the theory starts to break down for smaller system sizes at the local maxima and minima of the variance. This is expected since the theory is valid to order $N^{-1}$ in perturbation theory and the maxima and minima are where the effective local population is smallest.  Figure~\ref{fig:ringscale} b) shows the variance near a maximum as a function of $N$, showing an accurate prediction after $N=800$. The sample size for these microscopic simulations is $5\times 10^5$, and we estimate the error of the variance using bootstrap. The error is of order $10^{-2}$.
A segment of the spatio-temporal dynamics is shown in figure~\ref{fig:HeatMap}.  The theory matches the simulation quite well with the greatest deviation near the maxima and minima.
Figure~\ref{fig:multidimension} shows the two-time and two-space covariances of the synaptic drive for the same network parameters. The spatial covariance mirrors the coupling function as expected.

\begin{figure}[t!]
\begin{minipage}{.85\textwidth}
\includegraphics[width=1\linewidth, height=0.5\textheight]{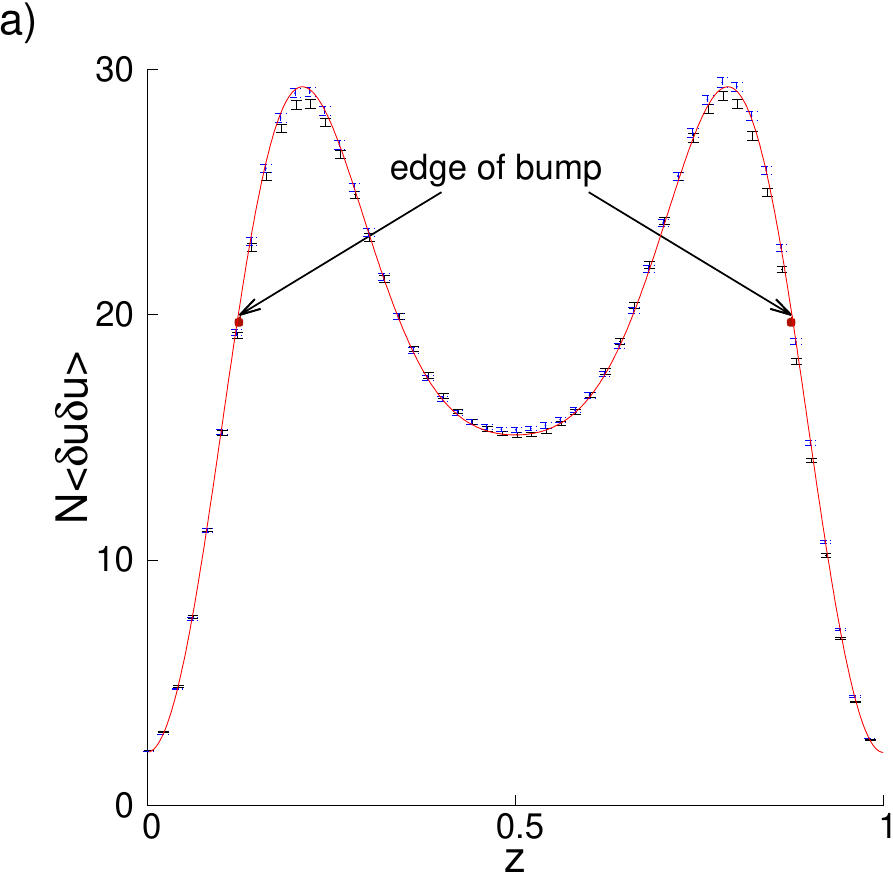}
\end{minipage}
\begin{minipage}{.45\textwidth}
\includegraphics[width=1\linewidth, height=0.3\textheight]{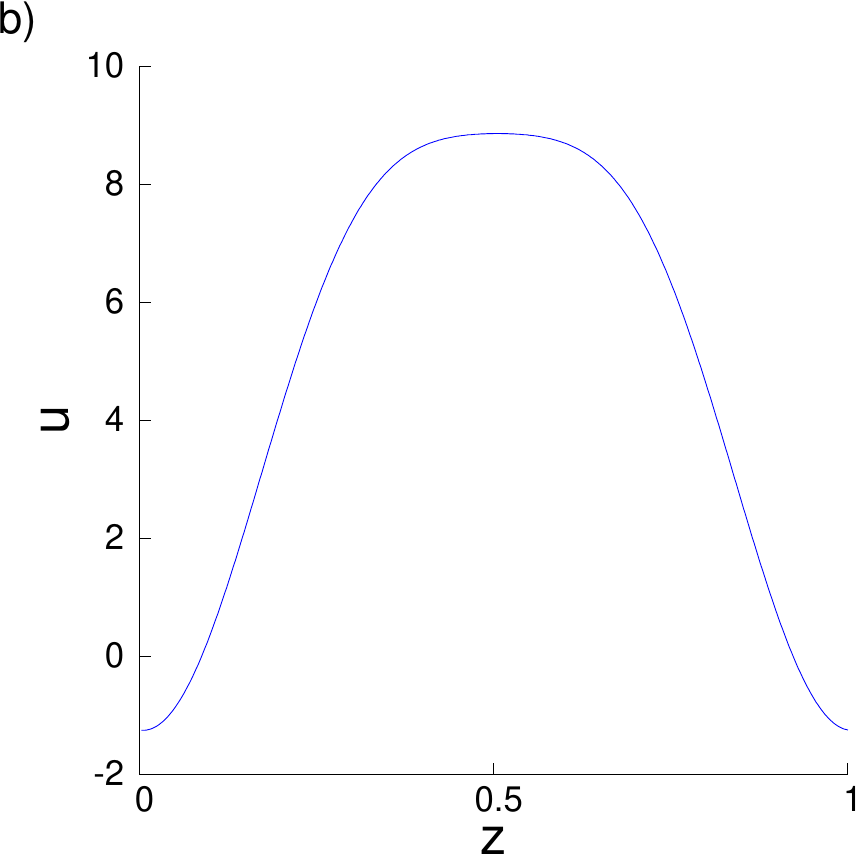}
\end{minipage}
\begin{minipage}{.45\textwidth}
\includegraphics[width=1\linewidth, height=0.3\textheight]{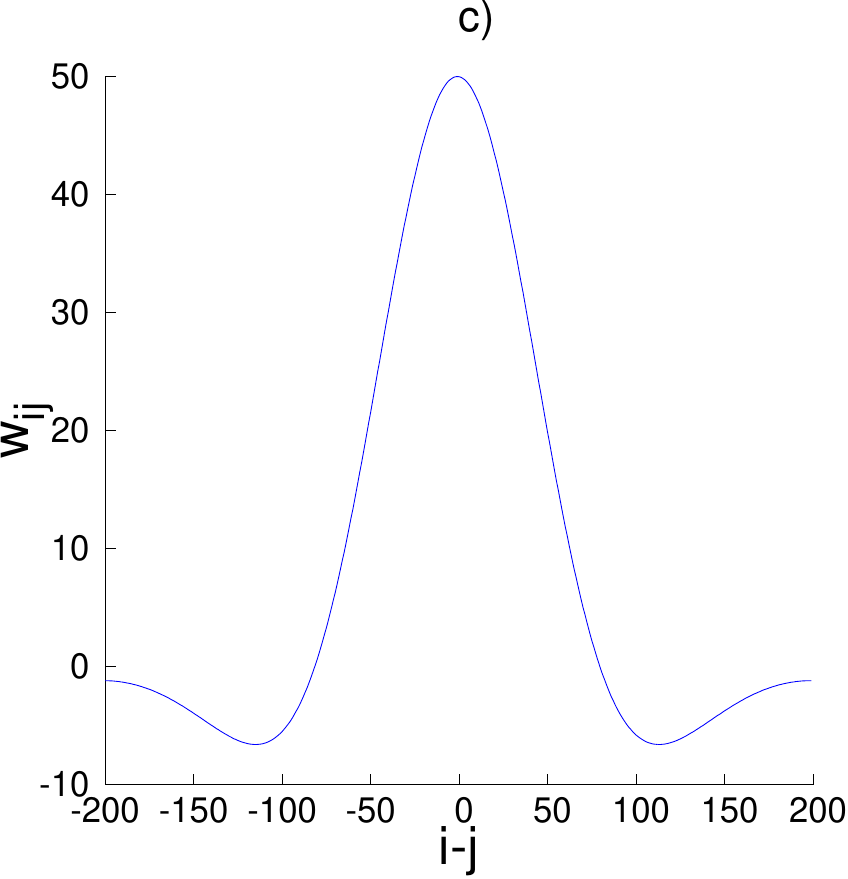}
\end{minipage}
\caption {a) Variance multiplied by $N$ and b) mean of the synaptic drive with subthreshold neurons for constant stimulus $I=-1$ and c) coupling weight  $w(z)=A\exp(-az)-\exp(-bz)+A\exp(-a(L-z))-\exp(-b(L-z))$ with $A=150$, $a=30$, and $b=20$. The supra-threshold edge of bump is at $u=1$. a) is evaluated at time 10. Standard errors are estimated by bootstrap.}
\label{fig:bumpmicroErrorBand}
\end{figure}

Figure~\ref{fig:bumpmicroErrorBand} shows a comparison between the theory and the microscopic simulation when subthreshold neurons are included. There is a good match when $N$ is large. As $N$ decreases the theory starts to fail at the edges of the bump first. This is likely due to the fact that the location of the edge could move and this is not captured by the theory since it assumes fluctuations around a stationary mean field solution. However, the spontaneous firing of sub-threshold neurons due to either the initial conditions or from the fluctuating inputs of supra-threshold neurons can cause the edge of the bump to move and this is a nonperturbative effect.

\subsubsection{Rate Covariance}
The firing rate is defined as $\nu = 2\eta(z,\pi,t)$
with mean 
\begin{align}
  \langle \nu(z,\pi,t) \rangle = 2 \rho(z,\pi,t)
\end{align}
and covariance
\begin{align}
\langle \delta \nu(z,t)\delta\nu(z',t')\rangle &= \langle(\nu(z,\pi,t) -\langle \nu(z,\pi,t)\rangle)(\nu(z',\pi,t')-\langle \nu(z',\pi,t')\rangle)\rangle\\
& = \langle \nu(z,\pi,t)  \nu(z',\pi,t')\rangle - \langle \nu(z,\pi,t)\rangle\langle  \nu(z',\pi,t')\rangle\\
&= 4\langle \eta(z,\pi,t) \eta(z',\pi,t') -4\rho(z,\pi,t)\rho(z',\pi,t')\rangle\nonumber\\
&=4 \langle (\tilde{\varphi}(z,t)  \varphi(z,t)+\varphi(z,t)) (\tilde{\varphi}(z',t')\varphi(z',t') +\varphi(z',t'))\rangle\\
&= 4\langle \varphi(z,t)\varphi(z',t')\rangle + 4 \langle \varphi(z,t)\tilde{\varphi}(z',t')\rangle\langle  \varphi(z',t') \rangle\\
&=4\langle  \varphi(z,t) \varphi(z',t')\rangle + \frac{4}{N}\Delta_\varphi^\varphi(z,\pi,t;z',\pi,t')\rho(z',\pi,t')
\end{align}
At tree level, from the diagrams in Figure \ref{fig:diagrams} b),
\begin{align}
N\langle  \varphi(z,t) \varphi(z',t')\rangle&=2\beta\int dz_1dz_2 d\tau \Delta_\varphi^v(z,t;z_1,\tau)\Delta_\varphi^\varphi(z',\pi,t';z_2,\pi,\tau)w(z_1-z_2)\rho(z_2,\pi,\tau)\nonumber\\
&+(x\leftrightarrow x')-\int dz_1\left\{\int d\theta\Delta_\varphi^{\varphi}(z,t;z_1,t_0,\theta)\rho(z_1,\theta,t_0)\right.\nonumber\\
&\times\left.\int d\theta'\Delta_\varphi^{\varphi}(z',t';z_1,t_0,\theta')\rho(z_1,\theta', t_0)\right\}
\label{eqn:ratecov}
\end{align}
We rewrite as
 \begin{align*}
N\langle  \varphi(z,t) \varphi(z',t')\rangle
&=2\beta\int dz_1dz_2 d\tau 
\left\{r^v(z,t;z_1,\tau)(r^{\varphi}(z',t';z_2,\pi,\tau)\right.\\
&\left.+\delta(\pi-\vartheta_>(\pi)+\nu_>(z')(t'-\tau))\delta(z'-z_2))
w(z_1-z_2)\rho(z_2,\pi,\tau)\right\}+(x\leftrightarrow x')\\
&-\int dz_1\left\{\int d\theta
(r^{\varphi}(z,t;z_1,\theta,t_0)+\delta(\pi-\vartheta_>(\theta)+\nu_>(z)(t-t_0))\delta(z-z_1))\rho(z_1,\theta,t_0)\right.\nonumber\\
&\times\left.\int d\theta'
(r^{\varphi}(z',t';z_1,\theta',t_0)+\delta(\pi-\vartheta_>(\theta')+\nu_>(z)(t'-t_0))\delta(z'-z_1))\rho(z_1,\theta', t_0)\right\}\\
&=2\beta\int  d\tau 
\int dz_1dz_2 r^v(z,t;z_1,\tau) r^{\varphi}(z',t';z_2,\pi,\tau)w(z_1-z_2)\rho(z_2,\pi,\tau)\\
&+\frac{2\beta}{|\nu_>(z')|}\sum_l\int dz_1r^v(z,t;z_1,t'-2\pi l/\nu)w(z_1-z')
\rho(z',\pi,\tau)\nonumber\\
&+(x\leftrightarrow x')-\int dz_1d\theta  r^{\varphi}(z,t;z_1,\theta,t_0)\rho(z_1,\theta,t_0)
\int d\theta'r^{\varphi}(z',t';z_1,\theta',t_0)\rho(z_1,\theta', t_0)\\
&-\frac{1}{2\pi}\int d\theta
r^{\varphi}(z,t;z',\theta,t_0)\rho(z',\theta,t_0)-\frac{1}{2\pi}\int d\theta'
r^{\varphi}(z',t';z,\theta',t_0)\rho(z,\theta', t_0)\\
&-\frac{1}{4\pi^2}\int dz_1\delta(z-z_1)\delta(z'-z_1)
\end{align*}
where
\begin{align*}
\Delta_\varphi^v(z,\pi,t;z',t')&=r^v(z,t;z',t')\\
\Delta_\varphi^\varphi(z,\pi,t;z',\theta',t')&=r^{\varphi}(z,t;z',\theta',t')+\delta(\pi-\vartheta_>(\theta')+\nu_>(z)(t-t'))\delta(z-z')\\
r(z,t;z',t')&=\int r^{\varphi}(z,t;z',\theta',t')\rho^0(z',\theta')\, d\theta'
\end{align*}

\begin{figure}[t!]
\begin{minipage}{0.9\textwidth}
\includegraphics[width=1\linewidth, height=0.15\textheight]{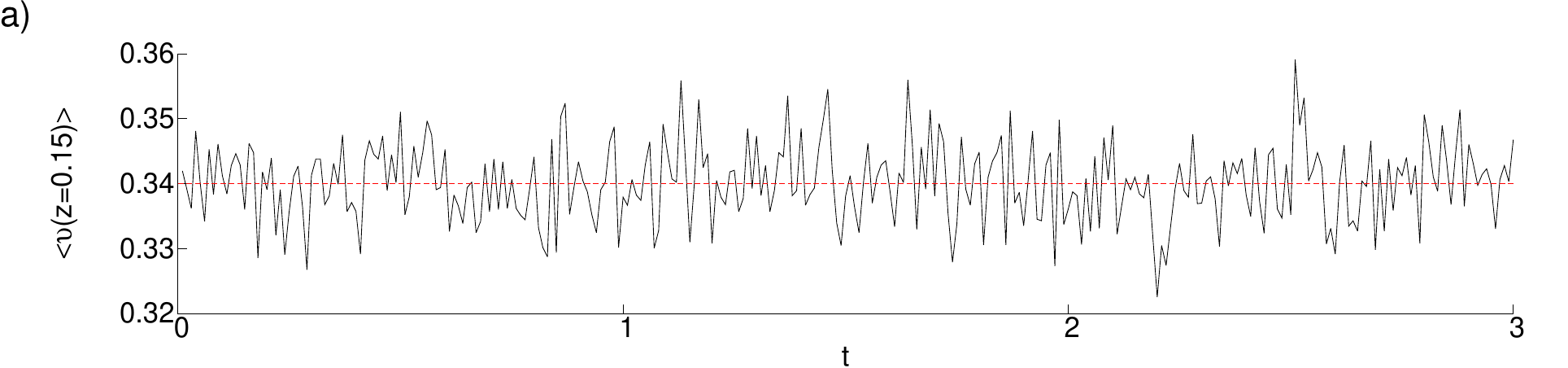}
\end{minipage}
\begin{minipage}{.4\textwidth}
\includegraphics[width=1\linewidth, height=0.2\textheight]{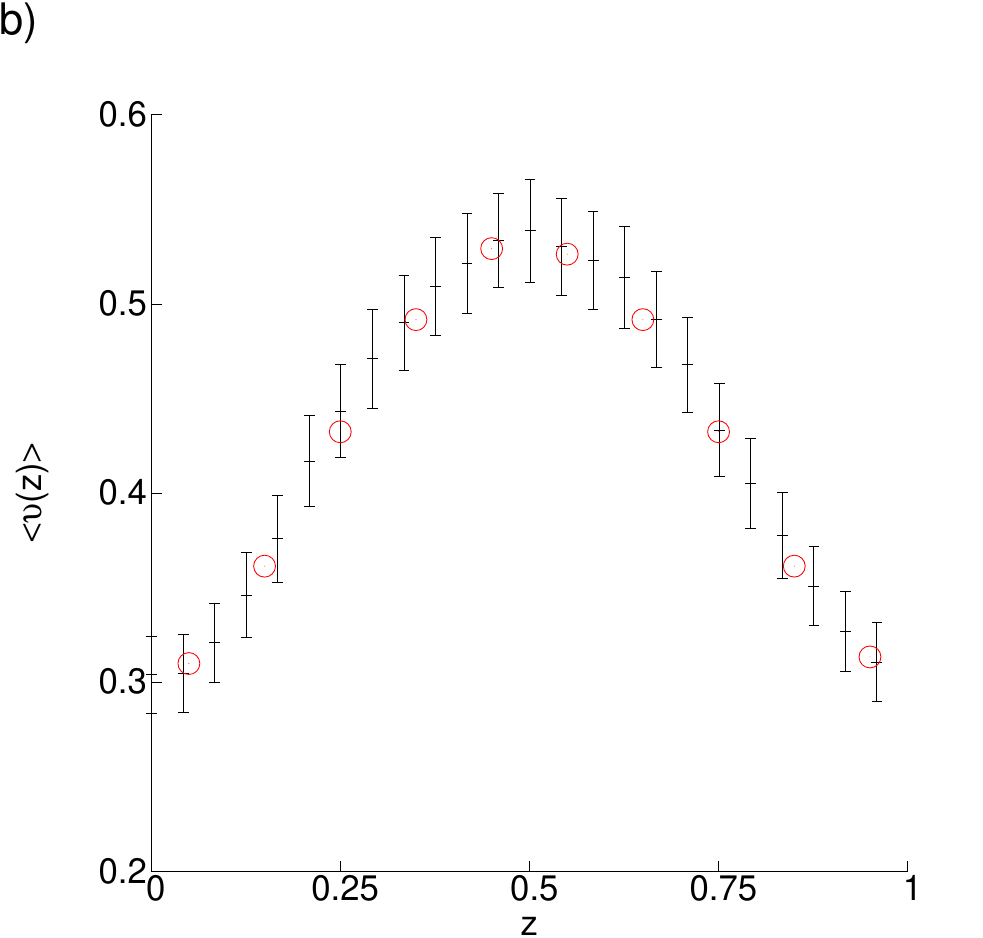}
\end{minipage}
\begin{minipage}{.9\textwidth}
\includegraphics[width=1\linewidth, height=0.15\textheight]{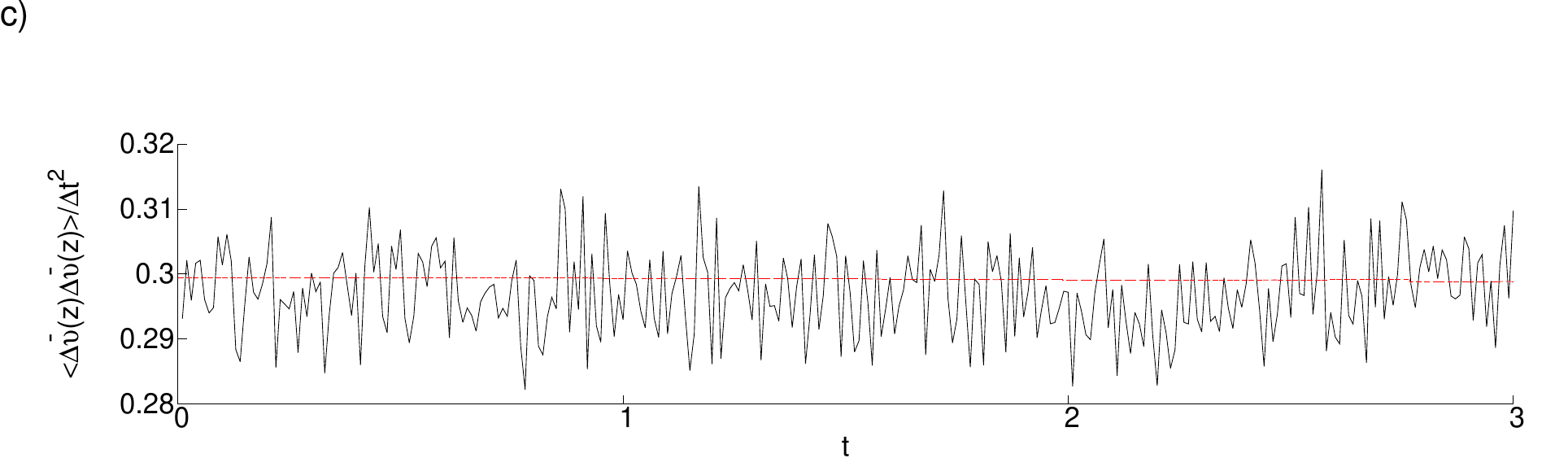}
\end{minipage}
\begin{minipage}{.4\textwidth}
\includegraphics[width=1\linewidth, height=0.2\textheight]{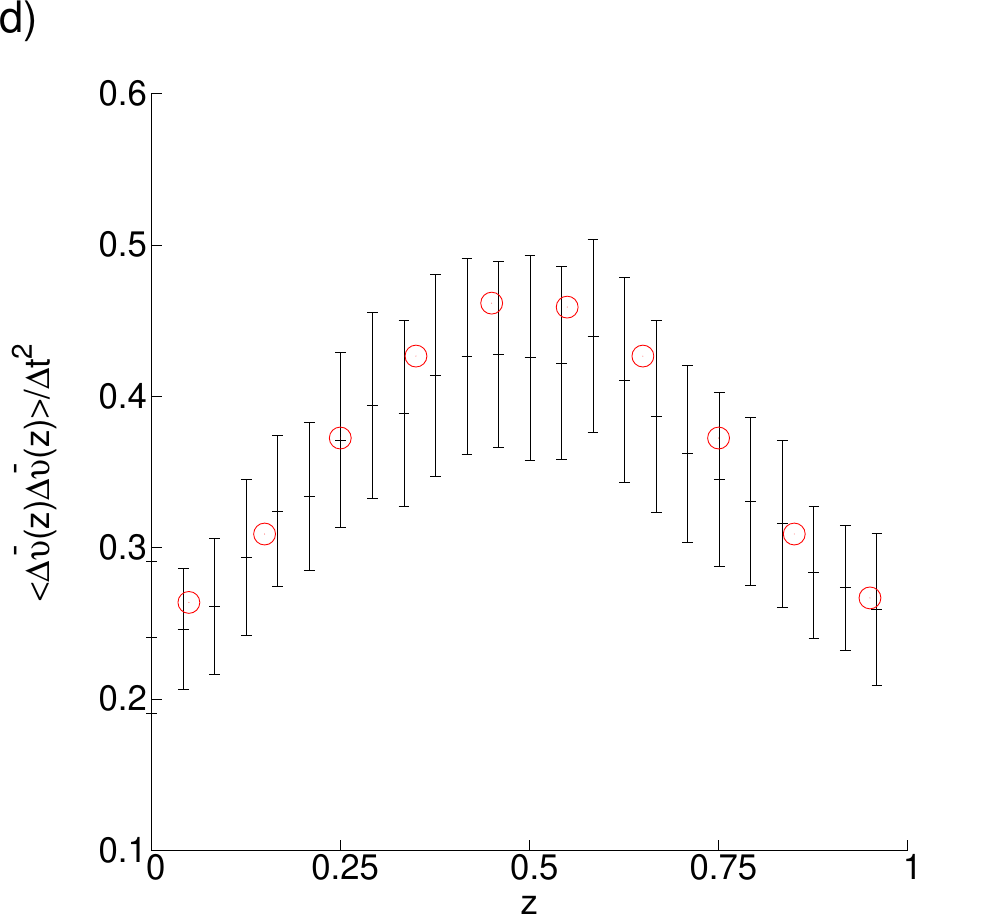}
\end{minipage}
\caption {a) Comparison between theory and microscopic simulations of time dependence of mean firing rate at one spatial location. b) Spatial dependence of mean firing rate at time 3.  c) and d) show the same comparisons for the variance given in equation (\ref{eqn:ratecov2}).
Parameters are from Fig~\ref{fig:ringsetup} and N=1200.
Standard errors are estimated by bootstrap. }
\label{fig:ratevariance}
\end{figure}

Hence
\begin{align*}
\langle \delta \nu(z,t)\delta\nu(z',t')\rangle  &=\nu_>(z)\nu_>(z')\bigg(\frac{8\beta}{N}\int  d\tau 
\int dz_1dz_2 r^v(z,t;z_1,\tau) r^{\varphi}(z',t';z_2,\pi,\tau)w(z_1-z_2)\rho(z_2,\pi,\tau)\\
&+\frac{8\beta}{|\nu_>(z')|N}\sum_l\int dz_1r^v(z,t;z_1,t'-2\pi l/\nu_>(z'))w(z_1-z')
\rho(z',\pi,\tau)+(x\leftrightarrow x')\\
&-\frac{4}{N}\int dz_1 r(z,t;z_1,t_0)r(z',t';z_1,t_0)\\
&-\frac{2}{\pi N}r(z,t;z',t_0)-\frac{2}{\pi N} r(z',t';z,t_0)-\frac{1}{\pi^2 N}\int dz_1\delta(z-z_1)\delta(z'-z_1)\bigg) \\
&+ \frac{4}{N}\Delta_\varphi^\varphi(z,\pi,t;z',\pi,t')\rho(z',\pi,t') 
\end{align*}
This quantity is well behaved for $t\ne t'$ and $z\ne z'$.  However, in
the limit of $t'\rightarrow t^-$, the rate covariance is singular since
\begin{align}
\lim_{t'\rightarrow t^-} \Delta_\varphi^\varphi(z,\pi,t;z',\pi,t')\rho(z',\pi,t')& =\delta(z-z') \delta(\nu_>(z)(t-t'))\left. \frac{d\phi}{d\theta}\right|_{\theta = \pi}\rho(z',\pi,t')\\
&= \frac{\nu_>(z)}{2\nu_>(z)}\delta(z-z')\delta(0)\rho(z',\pi,t')
\end{align}
We regularize the singularity at $t=t'$ by considering the time integral over a small interval:
\begin{align*}
\Delta \nu(z,t)=\int_{t-\Delta t/2}^{t+\Delta t/2}\delta\nu(z,s) ds
\end{align*}
giving
\begin{align}
\frac{\langle \Delta \nu(z,t)\Delta\nu(z',t')\rangle }{ \Delta t^2} &=\nu_>(z)\nu_>(z')\bigg(\frac{8\beta}{N}\int  d\tau 
\int dz_1dz_2 r^v(z,t;z_1,\tau) r^{\varphi}(z',t';z_2,\pi,\tau)w(z_1-z_2)\rho(z_2,\pi,\tau)\nonumber\\
&+\frac{8\beta}{|\nu_>(z')|N}\sum_l\int dz_1r^v(z,t;z_1,t'-2\pi l/\nu_>(z'))w(z_1-z')
\rho(z',\pi,t'-2\pi l/\nu_>(z'))\nonumber\\
&+(x\leftrightarrow x')-\frac{4}{N}\int dz_1 r(z,t;z_1,t_0)r(z',t';z_1,t_0)\nonumber\\
&-\frac{2}{\pi N}r(z,t;z',t_0)-\frac{2}{\pi N} r(z',t';z,t_0)-\frac{1}{\pi^2 N}\int dz_1\delta(z-z_1)\delta(z'-z_1)\bigg)\nonumber \\
&+ \frac{2}{N\Delta t}\rho(z',\pi,t') \delta(z-z') \nonumber\\
\label{eqn:ratecov1}
\end{align}
 We regularize the singularity at $z=z'$ by taking a local spatial average over $[-cN/2+z, cN/2+z]$. We make the approximation that within this local region, the propagator is constant on space, which is valid under the large $N$ limit. This results in
\begin{align}
\frac{\langle \overline{\Delta\nu}(z,t)\overline{\Delta\nu}(z',t')\rangle }{ \Delta t^2} &=\nu_>(z)\nu_>(z')\bigg(\frac{8\beta}{N}\int  d\tau
\int dz_1dz_2 r^v(z,t;z_1,\tau) r^{\varphi}(z',t';z_2,\pi,\tau)w(z_1-z_2)\rho(z_2,\pi,\tau)\nonumber\\
&+\frac{8\beta}{|\nu_>(z')|N}\sum_l\int dz_1r^v(z,t;z_1,t'-2\pi l/\nu_>(z'))w(z_1-z')
\rho(z',\pi,t'-2\pi l/\nu_>(z'))\nonumber\\
&+(x\leftrightarrow x')-\frac{4}{N}\int dz_1 r(z,t;z_1,t_0)r(z',t';z_1,t_0)\nonumber\\
&-\frac{2}{\pi N}r(z,t;z',t_0)-\frac{2}{\pi N} r(z',t';z,t_0)-\frac{1}{\pi^2 N c}\bigg)+ \frac{2}{\Delta t cN}\rho(z',\pi,t') \nonumber\\
\label{eqn:ratecov2}
\end{align}
Figure~\ref{fig:ratevariance} shows a comparison of the theory in (\ref{eqn:ratecov2}) to the microscopic simulations.  As shown in Fig.~\ref{fig:ratevariance} a), at $N=1200$, the theory predicts the mean firing rate well.  In Fig.~\ref{fig:ratevariance} c), we show the variance of the firing rate at fixed location. In Fig.~\ref{fig:ratevariance}d), we show the spatial structure of the variance.  Again, the theory captures the simulations.

\section{Discussion}

Our goal was to understand the dynamics of a large but finite network of deterministic synaptically coupled neurons with nonuniform coupling. In particular, we wanted to quantify the dynamics of individual neurons within the network. We first showed that a self-consistent local mean field theory can describe the dynamics of a single network if the external input and coupling weight are continuous functions.  This imposes a spatial metric on the network where neurons within a local neighborhood experience similar inputs and can thus be averaged over locally. This local continuity does not impose any conditions on long range interactions, which can still be random. We thus propose a new kind of network to study - {\it continuous randomly coupled  spiking networks}, where the coupling is continuous but irregular at longer scales.

We show that corrections to mean field theory can be computed as an expansion in the number of neurons in a local neighborhood. In this paper, we have chosen to scale the local neighborhood to the total number of neurons but this is not necessary. We do this by first writing down a formal and complete statistical description of the theory, mirroring the Klimontovich approach used in the kinetic theory of plasmas \cite{ichimaru1973basic, hora1984, 2007MichaelPRL}. This formal theory is regularized by averaging, which leads to a BBGKY moment hierarchy. As in previous works~\cite{2007MichaelPRE, MB2010, 2007MichaelPRL, finite2013, Michael2013bymf, MichaelGeneralized2013}, we showed that the Klimontovich description can be mapped to an equivalent Doi-Peliti-Jansen path integral description from which a perturbation expansion in terms of Feynman diagrams can be derived. The path integral formalism is a convenient tool for calculations.  Although we only computed covariances to first order (tree level) it is straight forward (although computationally intensive) to continue to higher order as well as compute higher order moments.  We only considered a deterministic network for clarity but our method can easily incorporate stochastic effects, which would just add a new vertex to the action. 

We showed that the theory works quite well for large enough network size, which can be quite small if all neurons receive suprathreshold input.  However, the expansion works less well for neurons with critical input such as neurons at the edge of a bump where infinitesimally small perturbations can produce qualitatively different behavior.  Quantitatively capturing the dynamics at the edge may require renormalization. The formalism could be a systematic means to understanding randomly connected networks~\cite{Sompolinsky1988} and 
the so-called balanced network \cite{Vreeswijk1996, Ostojic2014}, where the mean inputs from excitatory and inhibitory synapses are attracted to a fixed point near zero and the neuron dynamics is dominated by the fluctuations. 

\section{Acknowledgments}
This research was supported by the Intramural Research Program of the NIH, NIDDK.

\section{Appendix: Numerical Methods}
\subsubsection{Discretization schemes}
We use full backward's Euler for green function calculation for propagators.
\begin{align}
&\frac{dr_{ij}}{dt}= s_{ij}\\
&\frac{ds_{ij}}{dt}=\frac{1}{\pi} U_{ij}-\nu_i^2r_{ij}\\
&\partial_t   U_{ij} = -\beta U_{ij}+\frac{\beta}{N} \sum_j w_{ij}\nu_jr_{jk}+\frac{\beta}{2\pi}w_{ij}\nu_j
\end{align}
\begin{align}
&r_{ij}^t= r_{ij}^{t-1}+hs_{ij}^t\\
&s_{ij}^t=s_{ij}^{t-1}+h(\frac{1}{\pi} U_{ij}^t-\nu_i^2r_{ij}^t)\\
&  U_{ij}^t =   U_{ij}^{t-1} +h(-\beta U_{ij}^t+\frac{\beta}{N} \sum_j w_{ij}\nu_jr_{jk}^t+\frac{\beta}{2\pi}w_{ij}\nu_j)
\end{align}
\begin{align}
&r_{ij}^t-hs_{ij}^t= r_{ij}^{t-1}\\
&s_{ij}^t+h\nu_i^2r_{ij}^t-h\frac{1}{\pi} U_{ij}^t=s_{ij}^{t-1}\\
&  U_{ij}^t -h\frac{\beta}{N} \sum_j w_{ij}\nu_jr_{jk}^t+h\beta U_{ij}^t =   U_{ij}^{t-1} +h\frac{\beta}{2\pi}w_{ij}\nu_j
\end{align}
\begin{align}
\begin{pmatrix}
I & -hI & 0\\
h\nu^2.*I& I & - h/\pi I\\
-h\beta/N w.*\nu'&0 & I+h\beta I
\end{pmatrix}
\begin{pmatrix}
r_{\cdot j}^t\\
s_{\cdot j}^t\\
 U_{\cdot j}^t\\
\end{pmatrix}
=
\begin{pmatrix}
r_{\cdot j}^{t-1}\\
s_{\cdot j}^{t-1}\\
 U_{\cdot j}^{t-1} + h \beta/2\pi w_{.j}\nu_j\\
\end{pmatrix}
\end{align}


We add the spike terms in equation~(\ref{eqn:last}) directly to the propagator $\Delta_{\nu}^{\varphi}(z,t;z',\pi,t')$ for all possible $l$ when $t-T_l(z')=t'$. These spike terms add stiffness to the differential equation and explicit differential equation solvers like Runge-Kutta have poor stability properties.

\bibliography{Correlations-manuscript}
\bibliographystyle{plain}
\end{document}